\newif\ifShowKeys
\ShowKeysfalse

\documentclass[11pt,a4paper]{article} 
\pdfoutput=1
\usepackage[no-natbib-sort]{my-jheppub}

\usepackage{amsmath, amssymb}

\usepackage{mathpazo}
\usepackage{bm}
\usepackage{environ}
\usepackage{mathrsfs}
\usepackage{array,arydshln}
\usepackage{booktabs}

\usepackage{graphicx,epsfig}
\usepackage{epic}

\usepackage{tensor} 							

\usepackage{float}
\usepackage[dvipsnames]{xcolor}
\definecolor{maroon}{rgb}{0.8,0.3,0.}

\usepackage{slashed}
\usepackage[nodayofweek]{date time}

\ifShowKeys \usepackage[notcite]{showkeys} \fi

\usepackage{hyperref}

\usepackage{aurical}
\usepackage[T1]{fontenc}

\usepackage{framed}
\definecolor{shadecolor}{RGB}{255, 230, 204}

\allowdisplaybreaks

\newmuskip\pFqmuskip

\newcommand*\pFq[6][8]{%
  \begingroup 
  \pFqmuskip=#1mu\relax
  \mathcode`\,=\string"8000
  \begingroup\lccode`\~=`\,
  \lowercase{\endgroup\let~}\pFqcomma
  {}_{#2}F_{#3}{\left[\genfrac..{0pt}{}{#4}{#5};#6\right]}%
  \endgroup
}

\newcommand*\pFtildeq[6][8]{%
  \begingroup 
  \pFqmuskip=#1mu\relax
  \mathcode`\,=\string"8000
  \begingroup\lccode`\~=`\,
  \lowercase{\endgroup\let~}\pFqcomma
  {}_{#2}\widetilde{F}_{#3}{\left[\genfrac..{0pt}{}{#4}{#5};#6\right]}%
  \endgroup
}

\newcommand{\pFqcomma}{\mskip\pFqmuskip}

\newcommand{\be}{\begin{equation}}
\newcommand{\ee}{\end{equation}}

\newcommand{\mc}{\mathcal }

\newcommand{\bv}{\begin{verbatim}}
\newcommand{\ev}{\end{verbatim}}

\newcommand{\la}{\label}

\newcommand{\vp}{\varphi}

\def \ov {\over}
\def \ci {\cite}
\def \foot {\footnote}
\def \N {{\cal N}}
\def \b{\beta}
\def \m {\mu}
\def \n {\nu}
\def \del{\partial}
\def \p {\phi}
\def \ep{\epsilon}
\newcommand{\rf}[1]{(\ref{#1})}
\def \r {\rho}
\def \k {\kappa}
\def \l {\lambda}
\def \iffa {\iffalse}
\def \d {\partial} 
\def \repvec {\text{vector} }  
\def \repadj {\text{adjoint} }
\def \reptrip {\text{3-plet}}
\def \a  {\alpha}

\title{Partition function of free conformal fields\\ in 3-plet representation}

\author[a]{Matteo Beccaria} 
\author[b]{ and \ \ Arkady A. Tseytlin\footnote{Also at Lebedev Institute, Moscow.}}

\abstract{
Simplest examples  of AdS/CFT duality correspond to free CFTs   in $d$ dimensions  
with fields in vector or  adjoint representation of an  internal symmetry group 
  dual in the large $N$ limit to a theory of  massless  or  massless plus  
massive higher spins  in  AdS$_{d+1}$. 
One   may   also study    generalizations    when  conformal 
 fields belong to higher dimensional 
representations, i.e. carry more than two internal symmetry indices.
 Here we  consider   the  case of  the   3-fundamental   ("3-plet")  representation.  
One  motivation  is a   conjectured connection to 
multiple M5-brane   theory:  heuristic  arguments suggest  that it  may be  related to an  
(interacting)  CFT   of  6d  (2,0) tensor multiplets 
in 3-plet representation of   large $N$ symmetry group  that has  an AdS$_7$ dual.  
We compute the    singlet partition function $Z$ on $S^1 \times S^{d-1}$ 
 for a free field   in 3-plet representation of $U(N)$ 
  and  analyse  its  novel  large $N$    behaviour.
The large $N$ limit of the  low temperature   expansion   of $Z$ 
which  is  convergent  in the vector and adjoint cases  here 
is  only asymptotic,  reflecting the much  faster  growth of the  number  of  
singlet operators   with dimension,  indicating a phase transition at very low temperature.
 Indeed,   while  the critical  temperatures  in the vector ($T_c \sim N^\gamma ,\ \gamma >0)$ and adjoint ($T_c\sim 1$)
 cases are    finite,     
we find that in the 3-plet case  $T_c \sim (\log N)^{-1}$, i.e. it     approaches zero at large $N$. 
We discuss some  details of large $N$ solution for  the eigenvalue distribution.
 Similar  conclusions apply to higher $p$-plet  representations  of $U(N)$   or $O(N)$
      and also to the  free $p$-tensor   theories 
invariant  under $[U(N)]^p$  or  $[O(N)]^p$ with $p\geq 3$.  
}

\affiliation[a]{Dipartimento di Matematica e Fisica Ennio De Giorgi,\\
Universit\`a del Salento \& INFN, Via Arnesano, 73100 Lecce, 
Italy} 

\affiliation[b]{The Blackett Laboratory, Imperial College, London SW7 2AZ, U.K.}
                       
\emailAdd{matteo.beccaria@le.infn.it} \emailAdd{tseytlin@imperial.ac.uk}

\begin{document}

\def \g  {\gamma}\def \te {\textstyle} 
\def \zp {z_\Phi}


\begin{flushright}\small{Imperial-TP-AT-2017-{04}}\end{flushright}				

\maketitle
\flushbottom

\newpage
\section{Introduction}

\def \ie  {{\it i.e.} } \def \eg   {{\it e.g.}}

Examples of unitary conformal field  theories  are free  theories containing scalar, 
spinor or $({{d\ov 2} -1})$-form  fields 
in $d$ dimensions. Assuming  that these  fields  
transform in some representation    of a global $U(N)$ or $O(N)$ symmetry group   one expects 
  that in the large $N$ limit  this theory should be  dual  to a theory  in  AdS$_{d+1}$. 
  The latter should   contain 
  a tower of  massless   higher   spins   dual  to bilinear conserved currents
    as well as an  infinite   collection  of massive higher spins  
    dual to  primary  operators  containing irreducible ("single-trace")  contractions
    of more than two fields. 
    
The    simplest    case  is  that of  a fundamental  (vectorial) representation 
\ci{Klebanov:2002ja,Giombi:2016ejx, Giombi:2016pvg} 
where  the  spectrum   is particularly simple,   containing 
conserved currents in $CFT_d$   or massless   higher  spins in AdS$_{d+1}$.\foot{In the  $({{d\ov 2} -1})$-form  ({\em e.g.} 4d vector  field)   case there are also few additional massive  fields \ci{Beccaria:2014xda,Giombi:2016pvg}.}  
The   dual AdS   theory is then a   massless higher spin theory with 
 $N$ as the  coefficient in front of  the   classical action. 

Another   well-known    example   is  when  the  free  fields ({\em e.g.} spin-1 vectors in 4d)  
belong to the adjoint   representation
\ci{Sundborg:2000wp,HaggiMani:2000ru,Bae:2016rgm,Bae:2016hfy,Bae:2017spv}. Here  the dual AdS theory should be  string-like:
in  addition to massless higher spin fields  it
 should contain also families   of massive fields    which    together 
may have an interpretation  of  a spectrum 
of a   "tensionless"     string theory in 
AdS$_{d+1}$.\foot{In the maximally supersymmetric case this 
interpretation is of course suggested  by the  canonical example  of duality 
between the  $\N=4$  SYM theory  and the superstring theory in 
AdS$_5 \times $S$^5$  where  taking the 't Hooft  coupling to zero  corresponds  to a  
"tensionless" limit in the  quantum string theory.} 
The  coupling constant  on  the AdS   side    is then $1/N$, 
{\em i.e.}  the coefficient in front of the AdS  field  theory  action is  $N^2$. 
 
Here we would like to   study   the   next in  complexity   
case  when the  CFT field   belongs  to a  3-fundamental  or 
"3-plet"  representation,
  {\em  i.e. }  to  a general, symmetric   or antisymmetric 3-index tensor representation  
  of   a  global symmetry group. 
Already  in the  simplest  case of the  scalar  field   the spectrum   of   "single-trace" 
operators   with more than two fields 
(dual to massive  fields in AdS) 
 here is  much more   intricate  than 
in the adjoint case, suggestive of a  "tensionless membrane"  interpretation.  
The coefficient  in front of the    dual AdS  field theory  action  should now be $N^3$.

The spectrum of  CFT operators  on $\mathbb R^d$  (or states  in $\mathbb R \times S^{d-1}$) 
 is naturally encoded in 
the small  temperature $T=\beta^{-1}$  or small $x= e^{-\b}$  expansion  of the  partition function  $Z(x)$ 
 on $S^1_\b \times S^{d-1}$. 
 The  singlet-state  large $N$  partition function was previously computed in  the 
$U(N)$ \ci{Shenker:2011zf} 
and the $O(N)$  \cite{Jevicki:2014mfa,Giombi:2014yra} vector  representation case
and was matched to the corresponding 
massless higher spin partition  function in AdS. 
The  partition function in the adjoint representation  case  was  computed  in 
\ci{Sundborg:1999ue,Polyakov:2001af,Aharony:2003sx} 
(see also \ci{Skagerstam:1983gv,Schnitzer:2004qt,Barabanschikov:2005ri})  and its 
    matching  to  the  AdS  partition function was discussed  in
 \ci{Bae:2016rgm,Bae:2016hfy,Bae:2017spv}. 
Once the temperature  becomes large enough (of order  $N^\gamma$ with  
$\gamma >0$ in the vector case and $\gamma=0$ in the adjoint case) these  
partition functions exhibit 
phase transitions \ci{Shenker:2011zf,Sundborg:1999ue,Aharony:2003sx}  
 that  may have 
 some dual  AdS    interpretation ({\em cf.}      \ci{Amado:2016pgy}). 
 
Here  we  will  compute  the   singlet partition function  $Z(x)$ in the case of a  
free  CFT in a 3-plet ({\em i.e.} 
3-index tensor) representation of internal symmetry group.\foot{We shall consider the 
general  (reducible)   rank 3 tensor  $U(N)$ representation  as well as 
totally symmetric   and  antisymmetric 
 cases. We  shall also discuss   the  case of  the     3-plet representation of $O(N)$.
 The   large $N$  results will be similar in all of  these cases.}
We shall analyse     its  small temperature expansion   matching the direct operator count  
and  also present the large $N$ solution of the 
corresponding  matrix model  implying the presence of a  phase transition at  
small $x_c= e^{-\b_c}$  of order of $N^{-a}$ with $a> 0$, \ie at 
$T_c =  { a \ov \log N}$.

\subsection{Heuristic motivation:    3-plet 
(2,0)   tensor multiplet  as   M5-brane theory} 

Before   turning to the   details of    analysis  of the   large $N$   limit of the 
 3-plet partition   function  let us first make some speculative remarks 
 providing a 
    motivation  behind    this  work  which is  related  
to  attempts to understand a  6d CFT  describing     $N \gg 1$    coinciding 
 M5-branes   \ci{Witten:1995zh,Strominger:1995ac,kt1,kt2,Gubser:1997yh,Seiberg:1997ax,Gubser:1998nz,Henningson:1998gx,Harvey:1998bx,Aharony:1999ti, Bastianelli:1999ab,Bastianelli:2000hi,
 Tseytlin:2000sf,t1,Beccaria:2014qea,Beccaria:2015ypa}.\foot{Let us note  that   a possible  connection   of  multiple M5-brane  theory 
 to    interacting tensor models  was  mentioned in footnote 1 in \ci{Klebanov:2016xxf}. 
Interacting  tensor models  \ci{Gurau:2010ba,Gurau:2011xp,Carrozza:2015ad}  in large $N$   
limit  in  one   dimension   where  recently investigated 
 in \ci{Witten:2016iux,Klebanov:2016xxf} in connection with SYK model.  Some properties of 
scalar field    tensor models  in $d>1$  were   studied   in \ci{Klebanov:2016xxf}. 
 }
 
 The idea that  $N^3$  scaling of  observables  in the multiple  M5-brane case    may be explained  in terms of 
 M2-branes with three holes attached to three
different M5-branes   thus  leading  to 3-index  world volume fields   was  originally suggested  in \ci{kt2}.
Then a natural   proposal   that   a   6d superconformal 
  theory  describing multiple M5-branes    
  should involve  $(2,0)$ tensor multiplets 
  in 3-tensor  representation  of  $SU(N)$ or $SO(N)$  was  made in 
   \ci{Bastianelli:1999ab,t1}.\foot{Note that the   suggestion to  consider    tensor multiplets in a 3-plet representation   is different from  
  attempts  to construct  an interacting 
 theory of  tensor multiplets    assigned  to    adjoint representation  of  an internal  symmetry 
 group \cite{Samtleben:2011fj,Samtleben:2012mi}  or  using 3-algebra \ci{Lambert:2010wm,Lambert:2011gb}.}

To recall, the structure of a   world-volume theory   of a  single M5-brane   
  follows  from  
consideration of symmetries    and collective coordinates  corresponding to 
 the  classical M5-brane  solution in 11d supergravity  \ci{Gueven:1992hh,Callan:1991ky,Gibbons:1993sv,Kaplan:1995cp}.
 It is  represented by a free  $(2,0)$ tensor multiplet  containing 
(anti)selfdual 3-form $H_{\m\n\l}=3\del_{[\m} B_{\n\l]}$, 5 scalars $\p_r$ and 2 Weyl  fermions $\psi_a$. 
This is an example of a free 6d CFT with conformal anomalies and 
 correlators   that can be directly  computed  \cite{Bastianelli:1999ab,Bastianelli:2000hi}.
 
In   the case of  multiple   D-branes the low energy theory is  the SYM   theory, {\em i.e.} 
  one gets $N^2$  vector multiplets  at weak coupling  and that matches the   leading   $N^2$ scaling  predicted by   the  dual supergravity  for  BPS observables. 
  By   formal  analogy,  one   needs    $N^3$   free tensor multiplets to match the 
  leading    large $N$ scaling of  protected  11d  supergravity observables 
   \ci{kt1,kt2,Harvey:1998bx,Bastianelli:1999ab,Bastianelli:2000hi}.
   This suggests that  the  $(2,0)$   multiplet should   carry  a  3-index  representation of an     internal  
    symmetry group   that   has  dimension $\propto  N^3$  
    at large $N$.\foot{Explicitly,  the number of components 
    of  an antisymmetric 3-tensor is ${1\ov 6}  N(N-1) (N-2)$ and of a symmetric traceless 3-tensor is 
    ${1\ov 6}  N(N-1) (N+4)$.} 
 Thus,    if  the  conjectured   6d  theory  were to have a weak coupling limit 
 its    field content    would   be   $(B^{ijk}_{\m\n}, \p^{ijk}_r, \psi^{ijk}_a)$  with $i,j,k=1,2,..., N$.
    
  The reason why  the  M5-brane world-volume  2-form field     may 
  carry  3 internal  indices  can be  heuristically argued for as follows  
 \ci{kt2,Bastianelli:1999ab,kt1}. 
 Replacing the standard picture of  multiple D-branes connected   by open strings 
 by M5-branes  connected by M2-branes one  may  attempt  to 
 explain the $N^3$   scaling of  multiple M5-brane 
 entropy \ci{kt1} by assuming that triple  M5-brane connections   by "pants-like"  membrane  surfaces are 
 providing  the dominant contribution
  (pair-wise "cylinder" connections   should  give subleading $N^2$ contributions).\foot{The relevance of 
   triple M5-brane connections by membranes with 3 boundaries was suggested in \ci{kt2}  in
order to explain the $ \sqrt{N_1 N_2 N_3}$  scaling of the entropy of the 
extremal 4-d black hole described by the 2555 intersecting
M-brane configurations.  Similar virtual triple connections are 
not dominant  in the case of  open strings
ending on D-branes as  3-string interactions are subleading in  string coupling.
 3-hole "pants-like"   configurations  may be viewed as   basic building blocks  of 2d membrane   surfaces:
any membrane surface ending on several M5-branes may be cut into
"pants"   and  thus surfaces with more  than 3 holes   should   give  subleading  
contributions  in   membrane interaction strength. 
An indication   that 3-plet  M2-brane  interactions  may be  relevant  is  that    the 
$C_3$  form  of 11d supergravity naturally couples to  M2-brane world volume  while the 
 11d supergravity action contains  the cubic interaction term 
$C_3\wedge dC_3\wedge dC_3$.
In the limit when   M5-branes  coincide  and thus  the  M2-brane  configuration
 connecting them   becomes  small 
with string-like boundary,   the   membrane coupling  $\int C_3$  should lead to $\int B_2$    coupling at the
boundary  of a membrane   and this 2-form should  then 
carry $(ijk)$ indices. This is  analogous to how the 
$B_2$-field coupling to  world volume  of a  string translates into   vector $\int A^{ij} $ 
couplings at the two boundary points of an open string connecting two  D-branes.}

Alternatively, one  may think of  an  interacting  (2,0)  tensor multiplet theory as  a low-energy limit
of  a  tensionless  6d string theory  with  closed strings
 carrying  internal 3-plet  indices which  originate from virtual membranes connecting three
parallel M5-branes:  when the  M5-branes get close,   the membranes with 3 holes
may effectively   reduce to  strings that  then have  3 internal  labels  and  thus 
   $B^{ijk}_{\m\n}$  massless  modes. 
The   correlation   between the 3 Lorentz indices and the 3 internal indices of  the   
corresponding field strength 
$H^{ijk}_{\m\n\l} = d B^{ijk}_{\m\n}$   is thus   a  direct  analog   
of the   fact that  the   YM field  strength which is a rank 2  Lorentz tensor 
carries also  2  internal  indices (takes values in adjoint representation).

Ignoring  
  the self-duality constraint   on $H_{\m\n\l}$   one may  start  with  a free  theory 
\be \la{1.1} 
L_0= H^{ijk}_{\m\n\l}  H^{ijk}_{\m\n\l} \ , \ \ \ \   
 \ \ \ \ \ \  \ \ \ 
\delta  B^{ijk}_{\m\n} = \del_{[\m} \ep^{ijk}_{\n]}  \ , \ee
where $\ep^{ijk}_\m$ is a   gauge parameter.  
  A  speculative 
idea of how  to generalize this to an interacting theory 
 is  to assume that   the role   of the 
   would-be  structure  constants   here should be   played   by  the 
    generalized  field strength $\widehat H^{ijk}_{\m\n\l}$ itself, {\em i.e.}  
     that a non-linear generalization of 
the transformation rule for $B$  in \rf{1.1}   should    have the following structure
\be 
\la{1.2} 
\widehat  \delta B^{ijk}_{\m\n} = \del_{[\m} \ep^{ijk}_{\n]}  +  
 \widehat H^{ij'k'}_{\m\r\l} B^{i'jk'}_{\n\r}  \ep^{i'j'k}_{\l}  + ...\  , \qquad \qquad 
   \widehat H^{ijk}_{\m\n\l}= \del_{[\m }B^{ijk}_{\n\l]}  +   
   \widehat H^{ij'k'}_{\m\r\k} B^{i'jk'}_{\n\r}  B^{i'j'k}_{\l\k}  + ...\  , \ee
where dots stand for terms with other possible contractions of indices. 
The  full  non-linear field strength   $\widehat H$     should then be non-polynomial in $B$. 
Such  couplings   required to  contract 3-plet internal indices   may have a generalization to 
self-dual case   and may  correspond to a  
"soft"  gauge  algebra  structure  (thus possibly  avoiding no-go arguments  against the existence of an 
  interacting  chiral antisymmetric tensor theory in \ci{Bekaert:1999dp, Bekaert:1999ue}). 

One   consequence of  this   3-index  assumption     is that the leading  
 interaction   between the $B^{ijk}_{\m\n}$-fields 
should be {\it quartic}  \ci{t1}    rather   than    cubic as in the adjoint-representation YM  theory 
\be \la{1.3}
L= \widehat H^{ijk}_{\m\n\l}  \widehat H^{ijk}_{\m\n\l}\ \  \to\ \  \del B^{ijk} \del B^{ijk}  
  + B^{ijk}  B^{ij'k'} \del B^{i'jk} \del B^{i' j'k} + .... =
{\rm G}(B) \del B \del B +... \ .
\ee 
\iffa 
Then the  large $N$ limit may  be described  by iterated  "melonic" graphs as in tensor
 models 
 \ci{Gurau:2010ba,Gurau:2011xp,Carrozza:2015adg,Witten:2016iux,Klebanov:2016xxf}.
 \foot{In \ci{Gurau:2010ba} 
there was an attempt to use this fact to triangulate 3d  surfaces and that may have
 connection to membrane world surfaces then (see, however, remarks in \ci{Witten:2016iux}).}
  However, as clarified in \ci{Klebanov:2016xxf}   the melonic diagram description   of large $N$ limit 
 was proved   only     in the case when all 3  tensor   indices ${ijk}$  are distinguishable, {\em i.e.} 
 no symmetry is imposed and thus  one has $[O(N)]^3$ rather than 
 $O(N)$  symmetry  that is more appropriate to the  M5-brane theory case. 
 One  may still expect  such melonic representation to apply  but large $N$ expansion may be less  convergent 
 \ci{Klebanov:2016xxf}. 
 \fi
Adding scalar   fields of tensor multiplets  
 one may expect   to get similar  non-linear interactions, {\em e.g.}, through covariant derivative 
 $\ D_\m \p^{ijk} =  \d_\m \p^{ijk}   +   \widehat H^{ij'k'}_{\m\r\k} B^{i'jk'}_{\r\k}  \p^{i'j'k}+...$. 
Supersymmetry may require also  quartic and higher self-interactions of the 3-plet scalars   and  fermions.

Even assuming   
such a  hypothetical   interacting $(2,0)$ tensor multiplet  theory may exist  at the classical   level,  one
 faces several  difficult  questions.  The 
canonical  dimension of  the   free  $B$-field  in $6d$  is 2  and thus  $H$   has dimension 3. 
Then the classical interactions   in  \rf{1.2},\rf{1.3}  require   a dimensional  coupling parameter  
 and thus break  the classical conformal invariance of the free theory \rf{1.1}.   One   is then  to  hope 
 that at the quantum level there may exist a non-trivial conformal  fixed  point    at which  the dimension   of $B$  becomes 0.  
Another  important  question is about an   existence of  a well-defined large $N$ limit in such theory.\foot{In 3-tensor models 
with distinguishable indices ({\em i.e.} invariant under the $[O(N)]^3$  symmetry)
the  large $N$ limit may  be described  by iterated  "melonic" graphs 
 \ci{Gurau:2010ba,Gurau:2011xp,Witten:2016iux,Klebanov:2016xxf}.  
 As was noted in  \ci{Klebanov:2016xxf}, in the 
 case  when all 3  internal indices  of an interacting $\phi^4$ 
  scalar theory  are indistinguishable,   {\em i.e.}  transform, {\em e.g.},  in  symmetric representation 
 of a  single $O(N)$ group,    a   "melonic"  large $N$ limit  may still exist  but 
 the convergence of  the large $N$ expansion is unclear. 
 We thank  I. Klebanov and G. Tarnopolsky for  clarifying comments on this issue.} 
 The multiple M5-brane theory   should  certainly admit a  large $N$ expansion,  as 
 suggested, {\em e.g.},  by the  presence  of   11d  M-theory  corrections  to its anomalies
   \ci{Harvey:1998bx,
Tseytlin:2000sf, Beccaria:2014qea,Beccaria:2015ypa}  (see also \ci{Beem:2014kka,
Cordova:2015vwa}) 
   and its  free energy    \ci{Gubser:1998nz} 
which are subleading in $N$   compared to the leading  
$N^3$  term.     



As a starting point, one  may    consider  just   a free  3-plet  tensor multiplet 
 CFT    which  correctly describes  the $N^3$ 
 term in the anomalies   and    protected 3-point functions   as predicted by the 11d supergravity.  
  Regardless of  its   precise connection to   multiple M5-brane theory,    it 
  should   have a consistent AdS$_7$ dual on its own   right. 
 Our aim   below   will   be to   study the   thermal partition function
 in   such  free    tensor multiplet   theory  with fields in a  3-index  representation of 
an   internal  symmetry group.

\subsection{Structure  of the paper} 

Below we  shall   mostly concentrate on the $S^1 \times S^{d-1}$ 
partition function   a  free  scalar  field    in a 3-fundamental representation   of $U(N)$.   
The  cases of symmetric or antisymmetric   3-plet representation,  
 $O(N)$   symmetry  and 3-tensors  with distinguishable indices    will be  similar.  
The  generalization  to  free fermions or 
$({{d\ov 2} -1})$-form  fields    and thus,  in particular, to a (2,0)  tensor multiplet    will be straightforward. 

As we shall review in section \ref{sec2}, this 
 partition function  encodes  the spectrum of  "single-trace"  
primary operators in  the   free CFT. 
The singlet constraint   may  be implemented as in the  familiar  vector  or adjoint 
representation cases  by coupling  the 3-plet field
to a flat $U(N)$   connection and integrating over its non-trivial holonomy on $S^1$, \ie
   over a constant   matrix $U \in U(N)$.  For a free  field  $\Phi$ 
in a general real representation  $R$ of $U(N)$   the 
 resulting expression for  the partition function 
  $Z(x)$, \ $x= e^{-\b}$  will  be given by the  matrix $U$ integral  in \rf{2.5}
  with the   "action"  in the exponential  depending on  the   character $\chi_R (U)$ and   the 
  one-particle partition function $z_\Phi(x)$. 
  
  In section \ref{sec3}  we shall  compute the  low temperature  (small $x$)  expansion of $Z$ 
   in the large $N$ limit.  We shall first expand  the integral  \rf{2.5} in powers of $x$ 
   at finite $N$ and then  take $N \to \infty$. While in the   familiar cases of the 
    vector and adjoint  representations  the low temperature, $N=\infty$ 
      expansions  are    convergent (and thus   the $x\to 0$ and $N\to \infty$ limits
        commute in the  low temperature     phase  $T \leq T_c$), in the 3-plet case 
       this expansion will  be  only  asymptotic.  The reason  for this 
       will be a rapid growth of  the number of singlet operators with their dimension  which will lead to the  
       vanishing  of the  radius of convergence $x_c$  or the critical temperature $T_c$ 
      in the  $N=\infty$  limit.  
      We shall    find a closed  expression \rf{c6} for the $N=\infty$  limit of the 
        small $x$  expansion of $Z$ that  encodes  the number of  singlet operators  built out of elementary 3-plet 
fields  to any order in  dimension.
        
The  analysis  of  the phase structure of  $Z(x) $   at large  but finite $N$  will be carried out in \ref{sec4}.
  We  shall  rewrite the  integral representation for $Z$ in terms of  the   eigenvalue density 
  and study the large $N$ stationary-point solution for it. 
  We shall   find  that  in the 3-plet  case  there is a phase transition  at the critical temperature 
   $  T_{c}
   \sim \frac{1}{\log\,N}$  which approaches zero  at $N \to \infty$. 
At finite $N$ there are always two  phases, $ 0 < T \leq T_c$ and $T > T_c$, 
while at $N=\infty$  the  first phase  becomes  essentially the $T=0$  one,  
 so there is  only the  second  phase 
for any $T>0$.\foot{This is similar to   what happens  in the 1d Ising model  where for $T=0$ there   is no entropy term in the  free energy 
and ordered phase is favored.} 
As  in the  vector and adjoint cases,   we will find that  in the  $T >T_c$   phase the  large $N$ 
stationary point  solution for the   eigenvalue density leads to $\log Z = \mc O(N^2)$.

Some concluding remarks  about open issues    will be made in section \ref{sec5}. 
 Few  technical details     and extensions  will be presented  in   Appendices. 
In particular, the case of 2-plet    representation will be  discussed  in Appendix \ref{app:A}. 
The  $O(N)$ case   will be considered  in Appendix \ref{app:X}. 
In Appendix \ref{app:K}   we shall compute the value of  the eigenvalue density action on the 
stationary-point solution  in the 3-plet  case. 
In Appendix \ref{app:II}   we shall   give a general   relation   for  the "single-trace" partition   function 
counting only irreducible contractions  in terms of the full $Z$. 

In Appendix \ref{app:F}  we shall    analyze the  singlet partition  function 
of a $p$-tensor  with all  indices   being different, i.e. transforming  under a separate copy  of $U(N)$
as in the tensor models in \ci{Witten:2016iux,Klebanov:2016xxf}.  
It turns out that the  case of  such $[U(N)]^p$   theory is very similar to the $U(N)$ one discussed above, with $p=3$ 
being again the critical value when  the small  temperature  expansion becomes only asymptotic in the large $N$ limit, 
\ie   with   the critical temperature being again  $T_c (N \to \infty)  \to 0.$


\section{Partition function with  singlet constraint}
\la{sec2}

Given a CFT, one  may be interested 
(in   particular,  in the context of AdS/CFT duality)  
 in its  thermal   partition function  $Z$ with a  singlet constraint  (see, \eg,  
\cite{Skagerstam:1983gv,Sundborg:1999ue,Aharony:2003sx,Schnitzer:2004qt,
Schnitzer:2006xz,Shenker:2011zf}). 
We shall consider   a  free field $\Phi$ in $S^{1}_{\beta} \times S^{d-1}$
transforming in a  representation of the global symmetry group. 
The singlet projection may be implemented by 
coupling $\Phi$ to 
a  flat connection $A_{\mu }=U^{-1}\partial_{\mu}U$ 
and integrating over it. 
Only the constant part of  the $A_0$ component   cannot be gauged away 
because of the non-trivial holonomy along the thermal circle. The resulting partition function  $Z$ is then given  by the 1-loop determinant with $A_0$-dependent covariant derivatives   integrated  over the constant eigenvalues of $A_0$
  \ci{Aharony:2003sx,Shenker:2011zf}.\foot{See  also 
  section 3 in \ci{Beccaria:2014zma}   for a discussion  of the case   when $\Phi$ is  a  4d  Maxwell  field and $R$ is a  vector representation of $U(N)$   or $O(N)$ group.}
This  gives  an equivalent   result to the coherent-state  approach  of \cite{Skagerstam:1983gv,Sundborg:1999ue}.


\iffa The advantage of this formulation is that, in principle, it is 
applicable to the interacting theory. In the free case, an equivalent alternative construction is available,
based on purely group theoretical considerations. For our discussion, it will be an effective tool and we now  briefly review it. 
\fi

In general, the  partition function on $S^1_\beta \times S^{d-1}$ may be written  as 
\be
\la{2.1}
Z = \sum_{\text{singlets}}x^{E},\qquad\qquad\qquad   x = e^{-\beta}\ ,
\ee
where we assume  that the spatial sphere $S^{d-1}$ has a 
 unit radius  and $\beta = 1/T$. In a CFT,  the  discrete energy levels $E$ of states
on  
$S^{d-1}$ can  be identified with dimensions $\Delta$ of  the  corresponding 
operators
 in  $\mathbb R^{d}$. 
Before singlet projection, 
physical states are obtained by 
acting on the vacuum with suitable  composite operators built out of the 
elementary field $\Phi$. 
One   may define  the   {\em single-particle} partition function 
\be \la{222}
z_{\Phi}(x)=\sum_{i}x^{E_{i}},
\ee
 that counts  all such states treating $\Phi$ as a singlet,
{\em i.e.}  enumerates the components  of $\Phi$ and its derivative descendants 
modulo free equations of motion (thus  having also the interpretation of  the 
 character of the corresponding representation of the conformal group). 
 For example,  in the case of  a   free CFT   represented by  a scalar  or 
Weyl  (or Majorana) fermion  in  dimension $d$   and a
  vector  in 4d or a  self-dual   rank 2  tensor  in 6d  one finds  the well known expressions (see, {\em e.g.}, 
  \cite{Kutasov:2000td,Gibbons:2006ij})
  \def \rmS  {{\rm S}}  \def \rmF  {{\rm F}}  \def \rmV  {{\rm V}}  \def \rmT {{\rm T}}  
\begin{align}
& \hskip 2cm z_{\rmS, d}(x) = \frac{x^{\frac{d}{2}-1} (1+ x)}{(1-x)^{d-1}}\ ,\qquad\qquad
z_{\rmF,d}(x) = \frac{2^{\frac{d}{2}}\,x^{\frac{d-1}{2}}}{(1-x)^{d-1}}\ ,\la{2.2}\\
& \hskip 2cm  z_{\rmV,4}(x) = \frac{6\,x^{2}-2\,x^{3}}{(1-x)^{3}}\ , \qquad\ \ \ \ \ \ \ \ \ \ \ 
z_{\rmT,6}(x) = \frac{10\,x^{3}-5\,x^{4}+x^{5}}{(1-x)^{5}}\ . \qquad \qquad \la{202}
\end{align}
For   a multiplet of free  conformal fields one is  to combine  properly the   contributions to $Z$   coming  from 
 $z_\Phi$ for individual fields  (cf. \rf{2.5},\rf{fff}).\foot{For  example,  if  one  {\it formally} 
sums up  the  single-particle partition functions in \rf{2.2},\rf{202}  taking the  fermion contribution 
   with the plus  sign  one  finds   for the  4d  $\cal N$=4   Maxwell  multiplet  and 6d (2,0) 
tensor multiplet  \ci{Beccaria:2014qea}
\begin{align}
\notag
&
z_{{\cal N}=4} =  6\,z_{\rmS,4} +   4 \,z_{\rmF,4}+z_{\rmV,4} 
= \frac{ 6 x + 16\,x^{\frac{3}{2}}+ 12\,x^{2} - 2x^{3}\ , \ \ \ 
}{(1-x)^{3}}  \ , 
 \\
&
z_{\rm (2,0)}=  5\,z_{\rmS,6} +   2 \,z_{\rmF,6}+z_{\rmT,6} = \frac{
5\,x^{2}+  16\,x^{\frac{5}{2}}+15\,x^{3}-5\,x^{4}+x^{5}
}{(1-x)^{5}}  
 \ . \notag
\end{align} 
Note that these  combinations actually appear in the full partition function 
\rf{2.5}  only in one ($m=1$) of the terms  
as   the fermionic  contribution enters with the sign $(-1)^{m+1}$
depending on the term  in the infinite sum in the exponent.
}
\iffa
\foot{For  example,  if we {\it formally} 
sum up  the  single-particle partition functions  taking the  fermion contribution 
   with the minus sign we get  for the  $\cal N$=4   Maxwell  multiplet in 4d and (2,0) 
tensor multiplet in 6d:
\begin{align}
\notag
&z_{{\cal N}=4} =  6\,z^{(4)}_{\rm S} -   4 \,z^{(4)}_{\rm F}+z^{(4)}_{\rm V} 
= \frac{ 6 x - 16\,x^{\frac{3}{2}}+ 12\,x^{2} - 2x^{3}
}{(1-x)^{3}}  = {2 (1 + 3 y)\ov (1+ y )^2} \ , \ \ \  \qquad   y\equiv  x^{-1/2} = e^{\beta/2} \ , \\
&z_{\rm (2,0)}= \te  5\,z^{(6)}_{\rm S} -    \,z^{(6)}_{\rm F}+z^{(6)}_{\rm T} = \frac{
5\,x^{2}- 16\,x^{\frac{5}{2}}+15\,x^{3}-5\,x^{4}+x^{5}
}{(1-x)^{5}} =  {1 + 4y + 5 y^2 \ov (-1 + y) (1 + y)^5} \ . \notag
\end{align} 
Note that these  combinations do not actually appear in the full partition function 
\rf{2.5}   where the fermionic  contribution enters with sign $(-1)^{m+1}$
depending on the term  in the sum in the exponent.
}
\fi

Let us focus on the simplest case of a single bosonic field $\Phi$ 
transforming in a real  representation $R$ of $U(N)$.\foot{If $\Phi$ is complex, $R$ will be the 
representation acting on the real components.  For example, 
if  $\Phi_{i}$  transforms in the fundamental representation  of $U(N)$,
$\Phi'_{i}=U_{ij}\,\Phi_{j}$, 
then   $R=N\oplus \overline N$, i.e. 
if 
$\Phi_{i} = \alpha_{i}+i\,\beta_{i}$ and $U = A+i\,B$,  then 
{\tiny
$\notag
\begin{pmatrix}
\alpha_{i}' \\ \beta_{i}'
\end{pmatrix}
= \mathscr U 
\begin{pmatrix}
\alpha_{i} \\ \beta_{i} 
\end{pmatrix}
=
\begin{pmatrix}
A & -B \\ B & A
\end{pmatrix}
\begin{pmatrix}
\alpha_{i} \\ \beta_{i} 
\end{pmatrix}  $.
}
This is a reducible $2N$ dimensional representation that may be identified with 
$N\oplus\overline N$. Its character  is  
\be
\notag
\text{tr}\,  \mathscr U = 2\, \text{tr}\,  A  = \text{tr}(U+U^{*}) = \text{tr}(U)+\text{tr}(U^{-1}) = 
\chi_{N}(U) + \chi_{\overline N}(U) \ . 
\ee
}
The full partition function $Z$ 
may be expressed as a sum over the occupation numbers of all modes, with a Boltzmann 
factor $e^{-\beta E}$ for  the total energy $E$ and a numerical factor counting 
the number of singlets in the corresponding 
product of representations. This gives
\begin{align}\la{2.3}
Z &= \sum_{n_{1}\ge 0}x^{n_{1}\,E_{1}} \sum_{n_{2}\ge 0}x^{n_{2}\,E_{2}}\dots
 \{\#\ \text{of singlets in}\ \text{sym}^{n_{1}}(R)\otimes \text{sym}^{n_{2}}(R)\dots\} \ . 
\end{align}
The number of singlets is obtained by integrating over the global symmetry group
 with the invariant Haar measure $dU$ (normalized as  $\int dU=1$)
\be\la{2.4}
Z  = \int  dU\,\prod_{i}\sum_{n_{i}\ge 0}x^{n_{i}\,E_{i}}\,
\chi_{\text{sym}^{n_{i}}(R_{i})}(U).
\ee
Using the explicit form of the character $\chi$ of the symmetric power $\text{sym}^{n}(R)$  (see,
{\em e.g.},  eq.~(A.8) in 
\cite{Aharony:2003sx}),  we can then write the singlet  partition function as 
\cite{Skagerstam:1983gv,Sundborg:1999ue}
\be
\la{2.5}
Z = \int dU\,\exp\Big\{\sum_i \sum_{m=1}^{\infty}\frac{1}{m}x^{m\,E_{i}}\,\chi_{R}(U^{m})
\Big\} =  \int dU\,\exp\Big\{\sum_{m=1}^{\infty}\frac{1}{m}\,z_{\Phi}(x^{m})\,
\chi_{R}(U^{m})\Big\} \ . 
\ee
Here we assumed that $\Phi$ is a boson; in the mixed  boson (B) + fermion (F)  case 
we  need to do the  replacement 
\be \la{fff}
z_{\Phi} (x^m)\ \  \to \ \  z_{B}(x^m) + (-1)^{m+1}  z_{F}(x^m) \  .\ee
\def \te {\textstyle}
\def \N {{R_0}}
Below we shall consider mainly  the following   representations 
 $R$ (with    characters $\chi_R$) 
\be
\la{2.6}
\begin{array}{ccc}
\addlinespace\toprule
R & & \chi_{R} \\
\toprule
{\rm vector} : N\oplus \overline N & \phantom{\qquad} & \text{tr}(U)+\text{tr}(U^{-1}) \\
{\rm adjoint}:  && \text{tr}(U)\,\text{tr}(U^{-1})\\
\text{3-plet} : N^{\otimes 3}\oplus{\overline N}^{\otimes 3} && [\text{tr}(U)]^{3}+[\text{tr}(U^{-1})]^{3}\\
\bottomrule\addlinespace
\end{array}
\ee
 In general,   for the   $p$-plet   field   transforming in the product of $p$  fundamental representations
 when $R= N^{\otimes p}\oplus{\overline N}^{\otimes p} $
one finds 
\be\la{266}
\chi_{  N^{\otimes p}\oplus{\overline  N^{\otimes p}}  }(U) = \big[\text{tr}(U)\big]^{p}+\big[\text{tr}(U^{-1})\big]^{p} \ .\ee
One  may also be interested in the antisymmetric or symmetric tensor  representations. 
For example, in the case of the 3-plet  representation one finds 
\be
\la{2.7}\te 
\chi_{(\N\otimes \N\otimes \N)_{\text{(anti)sym}}}(U) = \frac{1}{6}\,\big[\chi_{\N}(U)\big]^{3}
\pm\frac{1}{2}\,\chi_{\N}(U)\,\chi_{\N}(U^{2})+\frac{1}{3}\,\chi_{\N}(U^{3}),
\ee
where $R_0$ is a  fundamental or anti-fundamental representation, and  sign   (-)+ 
applies to  (anti) symmetric case.

\section{$N= \infty $ limit  of  low  temperature expansion of $U(N)$ partition function} 
\la{sec3}

In the  $N=\infty$ limit  the counting of states  is  expected to simplify  
because singlets can be constructed  without considering special features of the finite $N$ case
\cite{Sundborg:1999ue,Polyakov:2001af}. 
In general,  $Z$    may be   expressed in terms of the  "single-trace"  partition function  
$Z_{\rm s.t.}(x)$  -- a generating 
 function   of 
 fully connected (indecomposable) 
contractions of fields 
\be\la{333}
\log Z(x)  \equiv \sum_{m=1}^{\infty}\frac{1}{m}\,Z_{\rm s.t.}(x^{m}) \ . 
\ee
The  expression for $Z_{\rm s.t.}(x)$  is   well known   
 in the  vector and
adjoint  representation cases  and  can   be  generalized   to   higher representations.  
As we  shall discuss  in Appendix \ref{app:II}, one can  formally invert the relation \rf{333} 
and determine $Z_{\rm s.t.}(x)$  in terms of  $Z(x)$. 

Below,  in  section \ref{s1},    we shall   review the known expressions for the $N=\infty$ partition functions
in the vector  and adjoint cases  and then in section \ref{s2}  turn to the 3-plet case.
We shall  first  consider the expansion of the 3-plet $Z$ in \rf{2.5}   in small $x$   for  finite $N$ 
and then  take the $N \to \infty$  limit of the coefficients at each order in $x$.  
In contrast to the  vector and adjoint  cases,  here the small $x$ and  large $N$ limits will not 
commute: 
the large $N$ limit of the  $x^n$ coefficients in  $Z$ will grow too  fast with $n$  so 
 that  the small $x$ expansion will not converge, {\em i.e.} the radius of convergence goes to 
  zero  in the $N\to \infty$ limit. 
  
 The reason for that will be understood  in   section \ref{sec4} where 
 will find that  there is a phase transition at  the critical temperature $T_c \sim (\log N)^{-1} $ 
 which goes to zero  when $N\to \infty$.
 Thus  the  low  temperature phase effectively disappears  (shrinks to $T=0$) in the strict $N=\infty$ limit.

In section \ref{SS}   we shall  present 
an exact expression that  summarizes the asymptotic    low temperature 
expansion of  the $N\to \infty $ 3-plet partition function.
This closed form     encodes  the number of  singlet operators  built out of elementary 3-plet 
fields  at  any order in  dimension.

\subsection{Vector and adjoint cases} 
\la{s1}

 In  the vector 
representation  case the singlets in $\text{sym}^{n}(N\oplus\overline N)$ in \rf{2.5} 
are products of 
invariant bilinears in  operators  built  out of  $\Phi$ and $\overline \Phi$.
For  example, 
 the bilinear singlets  built out of a scalar   field  have the 
form $\sum_{ss'}c_{ss'}\sum_{i}\partial^{s}\overline \Phi_{i}\partial^{s'}\Phi_{i}$.\foot{There are also singlets built using 
the  invariant $\epsilon_{i_1....i_N}$ tensor but their effect on the 
partition function  is  exponentially suppressed at large $N$   
 \cite{Shenker:2011zf}.} 
The 
partition function of such bilinears is  the square of the single-particle partition function  \rf{222}, \ie
the  "single-trace" partition function here is 
\be \la{3311}
Z^{\repvec}_{\rm s.t.}(x)=\big[z_{\Phi}(x)\big]^{2}  \ . \ee
 Including products of the bilinears, {\em i.e.} of 
all possible singlets, we get the $N=\infty$ partition function for  the vector representation
in the form \rf{333}       \cite{Shenker:2011zf}
\be
\la{3.1}
\log Z^{\repvec} = \sum_{m=1}^{\infty}\frac{1}{m}\,\big[z_{\Phi}(x^{m})\big]^{2} \ . 
\ee
In the adjoint case the singlets are built as products of single-trace operators.
The partition function for single-trace operators may be  found  by  the Polya 
enumeration theorem 
\cite{Sundborg:1999ue,Polyakov:2001af}
\be
\la{3.2}
Z_{\rm s.t.}^{\repadj} = -\sum_{m=1}^{\infty}\frac{\varphi(m)}{m}\,\log\big[1-z_{\Phi}(x^{m})\big]\ .
\ee
Here $\varphi(m)$ is the Euler's totient function counting  all  positive integers up 
to a given integer $m$ that are relatively prime to $m$.
The full partition function is obtained by considering multi-trace singlets treating single trace states as 
identical particles. As a result, the  $N=\infty$ partition function in  the 
adjoint case is given by 
\be
\la{3.3}
\log Z^{\repadj} = \sum_{m=1}^{\infty}\frac{1}{m}\,Z_{\rm s.t.}^{\repadj}(x^{m}) = -\sum_{m=1}^{\infty}
\log\big[1-z_{\Phi}(x^{m})\big]\  .
\ee
As was already mentioned   above (see  \rf{fff}),  in the   general  boson plus  fermion  field case  one has   to replace 
$z_{\Phi} (x^m) $ by $ z_{B}(x^m) + (-1)^{m +1}z_{F}(x^m) $ in \rf{3.1}   and \rf{3.3}.

From the AdS/CFT 
perspective, the bilinears in  the vector case are in  direct  correspondence with the 
massless  higher spin fields in AdS.  Hence  the  total partition function (\ref{3.1})  should 
match  the AdS   partition function  \ci{Shenker:2011zf,Giombi:2014yra,Beccaria:2014xda}.
Similarly, 
in the adjoint case, the  single trace operators  correspond to the spectrum
massless   and  massive  higher spin  fields  in
AdS. 
On  the  group-theoretical  basis,   one   expects again to    match   the  full
multi-particle partition function  (\ref{3.3}) with  its AdS  counterpart 
 \cite{Bae:2016rgm,Bae:2016hfy,Bae:2017spv}.

\subsection{3-plet case:    low temperature  expansion  of $Z$  and  counting of operators}
\la{s2}

For fields transforming in a  generic  representation  direct 
counting of states  may be very cumbersome.
One  may   work  out 
the direct 
   expansion  of   the partition function  (\ref{2.5})  in powers of $x=e^{-\b}$
   that effectively encodes  the counting of singlet operators. 
Taking into account the specific form of the characters (cf.  (\ref{2.6}))
one then  needs to compute   the $U(N)$ group   integrals 
\be
\la{3.4}
I(\bm a, \bm b) = \int dU\,\prod_{\ell\ge 1} (\text{tr}\,U^{\ell})^{a_{\ell}}\,
\overline{(\text{tr}\,U^{\ell})^{b_{\ell}}}, 
\ee
where $\bm{a}=(a_\ell)$ and $\bm{b}= (b_\ell)$   are sets of integers. 
Such  integrals can be found  straightforwardly 
 for  finite $N$, but the computational complexity 
grows rapidly  with increasing $N$. There are efficient  techniques to 
improve the computation, see for instance \cite{Balantekin:1983km,puchala2017symbolic}. 
Here  we are interested in the large $N$ limit. 
The integrals  (\ref{3.4}) 
are  zero if 
\be \la{344}
\kappa(\bm a)  \equiv  \sum_{\ell}\ell\,a_{\ell}  \neq  \sum_{\ell}\ell\,b_{\ell}\ee
 and
do not depend on $N$ as soon as $\kappa(\bm{a})\le N$, see 
  \cite{pastur2004moments}.
At each order in small $x$ expansion
we find a finite set of integrals $I(\bm a, \bm b)$ and the  above condition 
is certainly true if $N$ is sufficiently large, {\em i.e.}
 it is not actually necessary to  consider  the case of $N=\infty$ directly.
One then  finds the following  simple result
\be
\la{3.5}
I(\bm a, \bm b) = \prod_{\ell\ge 1}\ell^{a_{\ell}}\,a_{\ell}!\,\delta_{a_{\ell}, b_{\ell}}.
\ee
Let us consider, for example,  the case  when the field $\Phi$ is a 4d scalar  with 
 the one-particle partition  function $z_\Phi(x)=z_{\rmS,4}(x)$  given  in  \rf{2.2}.
If the scalar is in  the vector  
representation of $U(N)$,   expanding $Z$ in  \rf{2.5} in powers of $x\ll 1 $ 
 one finds 
\begin{align}
Z^{\repvec}_{\rmS,4} &= 1+\int dU\,\Big[x^{2}\,\tau_{1}+8\,x^{3}\,\tau_{1}+x^{4}\,(34\,\tau_{1}+
\tfrac{1}{4}\tau_{1}^{2}+\tfrac{1}{4}\,\tau_{2})+
x^{5}\,(104\,\tau_{1}+4\,\tau_{1}^{2})\notag \\
&\ \ \ \qquad\qquad +x^{6}\,(259\,\tau_{1}+33\,\tau_{1}^{2}+\tfrac{1}{36}\,
\tau_{1}^{3}+2\,\tau_{2}+\tfrac{1}{4}\,\tau_{1}\,\tau_{2}+\tfrac{1}{9}\,\tau_{3})+\dots\Big]\ ,\la{377}
\end{align}
where $\tau_{\ell}(U) \equiv (\text{tr}\,  U^{\ell})\,\overline{(\text{tr}\,  U^{\ell})}$. 
Using (\ref{3.5}) 
we  find  at $N\to \infty$ 
\be
\la{3.7}
Z^{\repvec}_{\rmS,4}  = 1+\,x^2+8 \,x^3+35 \,x^4+112 \,x^5+330 \,x^6+944 \,x^7+2849 \,x^8+\dots \ . 
\ee
The coefficient of a given power  $x^{k}$ stabilizes
 as soon as $N>N_{k}$ with $N_{k}$ growing  with $k$. 
Thus the expansion derived up to $x^{k}$ is exact as soon as $N>\max\{N_{k}\}$ which is 
a finite number. 
We  observe  that  \rf{3.7}  agrees with the small $x$ expansion 
of (\ref{3.1})  with $z_\Phi=z_{\rmS,4}$  in \rf{2.2}.
Note  that (\ref{3.1}) was  derived  directly at $N=\infty$.
 The reason for the agreement   is that here the $N\to\infty$ and $x\to 0$ limits  commute.
 This is reflected in convergence of the series in \rf{3.7}  which is also related 
 to the fact that the critical temperature  at which the low temperature   phase is no longer valid 
 here  goes to infinity at large $N$ (see  \rf{4.3x}).

Similarly, in the  adjoint case we  recover the expansion of (\ref{3.3}), {\em i.e.}
\be
\la{3.8}
Z^{\repadj}_{\rmS,4}  = 
1+\,x+6 \,x^2+20 \,x^3+75 \,x^4+252 \,x^5+914 \,x^6+3160 \,x^7+11194 \,x^8+\dots.
\ee
The series \rf{3.8}    have a finite radius of convergence   which reflects 
 the fact that  in the adjoint case the critical temperature  above which the 
low temperature  phase   no longer exists  is of order 1 (see \rf{4.4x}). 

{Let us also recall how the  direct counting  of operators goes in the vector and adjoint  $U(N)$ 
cases. For  the vector representation, 
the "single-trace"  (fully connected) 
 partition function in \rf{333},\rf{3.1}  is $[z_{\rmS,4}(x)]^{2}=x^{2}+8\,x^{3}+\dots$.
This  corresponds to one   operator  at  dimension 2 
$\overline\varphi_{i}\varphi_{i}$ and the $4+4$ operators 
$\overline\varphi_{i}\,\partial_{\mu}\,\varphi_{i}$ and 
$\partial_{\mu}\overline\varphi_{i}\,\varphi_{i}$  at dimension 4. 
In  the  adjoint case, the  single-trace
partition function  (\ref{3.2})  for a 4d scalar  is   $Z^\repadj_{\rm s.t.}=x+5\,x^{2}+\dots$. The  coefficients 
in this expansion   correspond to   one operator  $\text{tr}(\varphi)$ of  dimension 1 
and $1+4=5$ operators $\text{tr}(\varphi^{2})$  and $\partial_{\mu}\,\text{tr}(\varphi)$  of dimension 2.
}

\medskip
In the   novel   3-plet
representation  case  we  find from \rf{2.5},\rf{377}\rf{3.5} that   the 4d scalar 
 partition function  has the   following  large $N$ limit of  its  small  $x$ expansion (see  also 
 Appendix \ref{app:finiteN} for some finite $N$ data)
 \begin{align}
\la{3.9}
Z^{\reptrip}_{\rmS,4}  &= 
1+6 \,x^2+48 \,x^3+396 \,x^4+3504 \,x^5+35580 \,x^6+381216 \,x^7+4408956 \,x^8\notag \\
&\ \ \ \ \ \ \ \ +53647632 \,x^9+689785308 \,x^{10}+9258337104
   \,x^{11}+129842959752 \,x^{12}\notag \\
   &\ \ \ \ \  \ \ \  +1889221738416 \,x^{13}+\dots.
\end{align}
It is easy to see   how the  first three terms here may be  reproduced by   counting the  singlet 
operators.  
At dimension 2, we have  the singlets  built out of a scalar $\Phi=(\vp_{ijk})$  of   the structure
\be
\la{3.10}
(\overline \varphi\,\varphi) \ , 
\ee
i.e.  fully connected  contractions  $\overline\varphi_{ijk}
\varphi_{{i'j'k'}}$ where $i'j'k'$ is a permutation of $ijk$. 
This gives  $3!=6$  different cases. 
At dimension 3, we have the singlets
\be
\la{3.11}
(\overline \varphi\,\partial_{\mu}\varphi), \qquad (\partial_{\mu}\overline \varphi\,\varphi),
\ee
with the same contractions as above. This gives  total of $2\times 4\times 6 = 48$ 
of dimension 3 operators. At dimension 4 we get  the  bilinear structures
\be
\la{3.12}
(\overline \varphi\,\partial_{\mu}\partial_{\nu}\varphi), \qquad 
(\partial_{\mu}\partial_{\nu}\overline \varphi\,\varphi),\qquad
(\partial_{\mu}\overline\varphi\partial_{\nu}\varphi)\ . 
\ee
Ignoring terms vanishing on  the equation of motion  $\del^\mu\del_\mu \vp=0$
 gives 
$(9\times 2+4\times 4)\times 6 = 204$ operators. 
Then there is the  quartic  structure
\be
\la{3.13}
(\overline\varphi\varphi) (\overline\varphi\varphi)\ , 
\ee
where  we may have $6$ 
contractions in each factor;   
 counting  only once the symmetric cases,  we get    $\frac{1}{2}\times 6\times 7 = 21$ operators.
Finally,  there is also   an irreducible    "single-trace" operator  
\be
\la{3.14}
(\overline\varphi\varphi\overline\varphi\varphi) \ .
\ee
Here each $\overline\varphi\varphi$ appears in the form $\overline\varphi_{ijk}\varphi_{ijl} = X_{kl}$ where 
the free indices may be in any position and the two pairs of contractions may be in each of the two 
possibilities. This gives $3^{2}\times 2 = 18$ ways of constructing $X$. Contracting 
 the two $X$'s
(counting once the symmetric cases) gives  $\frac{1}{2}\times 18\times 19 = 171$
operators. 
The  total  number of dimension 4  singlet operators 
is then $204+21+171=396$,  in agreement with  the coefficient of the $x^4$ term in 
(\ref{3.9}).\footnote{
\la{ft13}
Let us note  that if we  formally replace 
 $z_\Phi(x)$ by $x$ in (\ref{2.5}),   that will correspond  to  considering  a {\em constant}
 complex   4d scalar  field 
 that has    dimension 1  (cf.   $z_{\rmS,4}$ in \rf{2.2}). 
 In this case $Z$  will count  just the $U(N)$ contractions leading to singlets. This counting 
  has been
worked out in \cite{Geloun:2013kta}. In the 3-plet representation case one  obtains
 the series 
$Z=1+6\, x^2+192\, x^4+10170\, x^6+834612\, x^8+90939630 \, x^{10}+12360636540\, x^{12}+\dots, $ which agrees with eq.~(86) of \cite{Geloun:2013kta}. The $192\,x^{4}$ term comes from the sum of 
the 21 structures in (\ref{3.13}) and the 171 in (\ref{3.14})   that do not involve derivatives.
}

 We may also  find the   corresponding "single-trace" generating function of 
 fully connected  contractions  
  (that starts with $x^2$ term) 
   by writing (\ref{3.9}) in the form \rf{333} 
\begin{align}
Z_{{\rm s.t.}}^\reptrip &= 
 6 \,x^2+48 \,x^3+375 \,x^4+3216 \,x^5+32098 \,x^6+342912 \,x^7+3976443 \,x^8\notag\\
&\quad  +48645632 \,x^9+629746974 \,x^{10}+8512245744
   \,x^{11}\notag \\
   &\quad +120220813597 \,x^{12}+1760740453968 \,x^{13}+\dots\ . \la{318}
\end{align}
The partition function  (\ref{3.9})  corresponds to a 4d scalar in  the general    3-index
tensor representation  of $U(N)$   without  any symmetry.
 In the case  of totally  symmetric $(+)$ or antisymmetric $(-)$  representations 
  we find, using the expressions for the  characters  in  (\ref{2.7})
\begin{align}
\la{3.17}
Z^{\reptrip^{+}}_{\rmS,4} &= 1+\,x^2+8 \,x^3+36 \,x^4+120 \,x^5+404 \,x^6+1368 \,x^7+5034 \,x^8
\notag \\
& \qquad +18736 \,x^9+71452 \,x^{10}+276864
   \,x^{11} +\dots, \\
\la{3.18}
Z^{\reptrip^{-}}_{\rmS,4}&= 1+\,x^2+8 \,x^3+36 \,x^4+120 \,x^5+403 \,x^6+1360 \,x^7+4978 \,x^8
\notag \\
& \qquad +18400 \,x^9+69645 \,x^{10}+267728
   \,x^{11}+\dots.
\end{align}
Compared to  (\ref{3.9}) we see that there there are fewer  operators  as some of the 
contractions are now equivalent.  Note that 
the  partition functions \rf{3.17} and \rf{3.18} 
start to  differ at order $\mc O(x^{6})$.
The  coefficients of the 
 first few  terms in (\ref{3.17}) or (\ref{3.18})  can be   easily  reproduced by the operator counting.\foot{The terms $x^{2}+8x^{3}$
are the same as  as in the  vector case (\ref{3.7}) as the correspond to 
 the operators (\ref{3.10}) and (\ref{3.11}) without  additional multiplicity as  the 3-indices are  now contracted in a unique way
($\overline\varphi_{ijk}\varphi_{ijk}$, etc.). 
The  coefficient  36 of the $x^{4}$  term  corresponds to  34  dimension 4 operators  in (\ref{3.12})
(now  with  unique contraction), one  operator as in (\ref{3.13})  of the  form 
$(\overline\varphi_{ijk}\varphi_{ijk})^{2}$, and one  additional  operator  
$\overline\varphi_{ijk}\varphi_{ijl}\overline\varphi_{pql}\varphi_{pqk}$  as in (\ref{3.14}).
For completeness,   let us list also  the "single-trace"  partition functions  corresponding to \rf{3.17}, \rf{3.18}:
\begin{align}
\notag
Z_{\rm s.t.}^{\reptrip^{+}}&=  \,x^2+8 \,x^3+35 \,x^4+112 \,x^5+332 \,x^6+968 \,x^7+3104 \,x^8\notag +10672 \,x^9+39466 \,x^{10}+153160
   \,x^{11}+\dots, \\
\notag
Z_{\rm s.t.}^{\reptrip^{-}} &= \,x^2+8 \,x^3+35 \,x^4+112 \,x^5+331 \,x^6+960 \,x^7+3049 \,x^8 +10352 \,x^9+37814 \,x^{10}+145192
   \,x^{11}+\dots.
\end{align}
}

Similarly, in the case of a scalar in the 3-plet representations  in 6 dimensions we find 
the following  analogs  of \rf{3.9},\rf{3.17},\rf{3.18} 
\begin{align}
\la{3.21}
Z^\reptrip_{\rmS,6} &= 1+6 \,x^4+72 \,x^5+456 \,x^6+2040 \,x^7+7452 \,x^8+26232 \,x^9\notag\\
&\qquad +111768 \,x^{10}+591432 \,x^{11}+3268332
   \,x^{12}+16860144 \,x^{13}+\dots, \\
\la{3.22}
Z^{\reptrip^{+}}_{\rmS,6} &= 1+\,x^4+12 \,x^5+76 \,x^6+340 \,x^7+1212 \,x^8+3676 \,x^9\notag\\
&\qquad+10032 \,x^{10}+25956 \,x^{11}+68632 \,x^{12}+196788
   \,x^{13}+\dots, \\
\la{3.23}
Z^{\reptrip^{-}}_{\rmS,6} &= 1+\,x^4+12 \,x^5+76 \,x^6+340 \,x^7+1212 \,x^8+3676 \,x^9\notag \\
&\qquad +10032 \,x^{10}+25956 \,x^{11}+68631 \,x^{12}+196776
   \,x^{13}+\dots.
\end{align}
As a 6d scalar has dimension 2,   the  low temperature   expansion  here  starts at $x^4$ order.
 The (anti) symmetric  3-plet partition functions \rf{3.22}, \rf{3.23} 
 differ at $\mc O(x^{12})$, {\em i.e.} at the level 
of operators with three $\overline\varphi\varphi$ pairs;  
this is in parallel   with what was  in 4d  where  the scalar dimension was 1 
(eqs.  \rf{3.17},\rf{3.18} differ   at $x^6$ order).  

Comparing  the coefficients  in  the vector \rf{3.7}, adjoint \rf{3.8} and 3-plet  cases \rf{3.9} 
we conclude that the number of  singlet  
operators  in the 3-plet case  grows much faster with 
the power of $x$, {\em i.e.}  with the  operator dimension.
This   implies   non-convergence   of the  small $x$ expansion.
Indeed, as we shall  find  in   section \ref{sec4} 
   an  analog  of the "Hagedorn"  transition  found 
 in the adjoint case \ci{Sundborg:2000wp,Aharony:2003sx}  
   here   happens at much lower temperature  $T_c \sim  (\log N)^{-1}  $ 
   which goes  to zero at $N \to \infty$.
This  will also become clear  from the closed expression  for the $N=\infty$  limit of the 
small $x$  partition function 
presented in the   subsection \ref{SS}. 

\def \zp {z_\Phi}

One 
   may   also  find similar low temperature expansions  of   $Z$ 
 in the case  when  the   global symmetry group  is  $O(N)$  instead of   $U(N)$; this is discussed in 
 Appendix  \ref{app:X}.
  
 \subsection{Exact  expression  for   low temperature  expansion of 
 3-plet   partition function in the $N\to \infty$  limit}
 \la{SS}
 
 Given a particular 
  choice of the character $\chi_{R}$ in  (\ref{2.5}) it is possible to 
 find  a closed form for the  small $x$  expansion of the 
 partition function  extending the   expansions like \rf{3.9}  to all orders in $x$. 
 
 Let us start  from  the following general form of  the series expansion of \rf{2.5} \footnote{We exploit the fact that the
  integral over $U$  vanishes   when  evaluated on   products of  characters 
of $U^{m}$ with different values of $m$. }
\be
 \la{mb1}
 Z =  \prod_{m=1}^{\infty}\sum_{k=0}^{\infty}\frac{1}{k!}\,\Big(\frac{z_{\Phi}(x^{m})}{m}\Big)^{k}\,\int dU\,\big[\chi_{R}(U^{m})\big]^{k} \ . 
\ee
If we consider the $p$-plet 
  (or  $N^{\otimes p}$ representation  of $U(N)$, then 
the corresponding character is given by \rf{266}, i.e. 
$\chi_{R}(U) = \big[\text{tr}(U)\big]^{p}+\big[\text{tr}(U^{-1})\big]^{p}$,  and the only 
non-vanishing contribution to  (\ref{mb1}) comes from the term with an equal number of 
$\text{tr}(U)$ and $\text{tr}(U^{-1})$ factors. Hence, using (\ref{3.5}), we find
\begin{align}
\la{mb2}
Z^{p\text{-plet}} &= \prod_{m=1}^{\infty}\mathop{\sum_{k=0}}_{k\ \text{even}}^{\infty}
\frac{1}{k!}\,\Big(\frac{z_{\Phi}(x^{m})}{m}
\Big)^{k}\binom{k}{k/2}\,m^{p\,k/2}\,(p\,k/2)!  \notag\\
&= \prod_{m=1}^{\infty}\mathop{\sum_{k=0}}^{\infty}
\frac{1}{(2\,k)!}\,\Big(\frac{z_{\Phi}(x^{m})}{m}
\Big)^{2\,k}\binom{2\,k}{k}\,m^{p\,k}\,(p\,k)! \notag \\
&= \prod_{m=1}^{\infty}F_{p}\big(m^{p-2}\,[\zp(x^{m})]^{2}\big),
\end{align}
where we have introduced the formal power series
 \be\la{c1}
 F_{p}(y) \equiv  \sum_{k=0}^{\infty} b_k \,y^{k}\ , \qquad \qquad b_k= \frac{(p\,k)!}{(k!)^{2}} \ , \qquad p=1,2,3,... \ . 
\ee
As a check, for $p=1$  and $p=2$ we find from \rf{c1}:
 \be 
\la{c3}
F_{1}(y) = e^{y}  \ , \ \ \ \ \ \qquad  \ \ \ \ F_{2}(y) = { 1 \ov \sqrt  {1-4y}}  \ , 
\ee
 and thus \rf{mb2} gives
 \begin{align}
 &\log Z^{\text{1-plet}} = \sum_{m=1}^\infty {1\ov m}  \big[\zp(x^{m})\big]^{2}  \ ,\la{c4} \\
& \log Z^{\text{2-plet}} = -\frac{1}{2}\sum_{m=1}^\infty\log\big(1-4\,\big[\zp(x^{m})\big]^{2}\big)\ , \la{c5}
 \end{align}
 which are indeed   the expressions for the  partition functions of the  vector  \rf{3.1}
 and 2-plet  \rf{B.3}   representations (cf. also \rf{E.8}).  
Let us mention for completeness  that in  the adjoint case  one finds,  similarly to \rf{mb2}, 
\begin{align}
Z^{\repadj} &= \prod_{m=1}^{\infty}\mathop{\sum_{k=0}}^{\infty}
\frac{1}{k!}\,\Big(\frac{z_{\Phi}(x^{m})}{m}
\Big)^{k}\,m^{k}\,k! = \prod_{m=1}^{\infty}\sum_{k=0}^{\infty}\big[z_{\Phi}(x^{m})\big]^{k} = 
\prod_{m=1}^{\infty}\big[1-z_{\Phi}(x^{m})\big]^{-1},
\end{align}
which is  indeed  the partition function in  (\ref{3.3}).

Surprisingly, 
the series $F_p$ in \rf{c1}    no longer converges (has  zero radius of convergence)\foot{This is implied by the growth of 
  ${b_{k+1}\ov b_k}= {(kp + p)!\ov (k+1)^2 (k p)! }\Big|_{p=3} =  27 k +  { 6\ov 1 + k}$, with $b_k$   defined in \rf{c1}.}
starting with  $p=3$.   
Thus for $p\geq 3$  \rf{c1} should be regarded only as a formal generating function. 
One may  "resum"   \rf{c1} 
by first 
replacing the numerator $(3\,k)!$ in $b_k$ in \rf{c1}   by $ \int_{0}^{\infty} dt\,e^{-t}\,t^{3\,k}$, then 
summing over $k$  and finally   integrating over  $t$.
In this way  we obtain 
\be\la{cc5}  F_3(y) \ \to \ 
\widetilde F_{3}(y) = \tfrac{1}{6}\,\sqrt{\tfrac{\pi}{3}\,y^{-1}}\ e^{-\frac{1}{54}\,y^{-1}}\,\Big[
I_{1\ov 6}\big(\tfrac{1}{54}\,y^{-1}\big)+I_{-{1\ov 6}}\big(\tfrac{1}{54}\,y^{-1}\big)\Big],
\ee
where $I_{a}(y)$ is the 
modified Bessel function of the first kind.
The function   $\widetilde F_{3}(y) $ has a branch cut
on the  negative real  axis, but is real, positive and smooth for $y\ge 0$. The power series \rf{c1}
defining $F_{3}(y)$  is precisely an 
{\em asymptotic} expansion of $\widetilde F_{3}(y) $   for $y\to 0^{+}$, 
i.e. 
for any  integer $K$  and  $y\to 0^{+}$
  we have $\widetilde F_{3}(y)-\sum_{k=0}^{K}\frac{(3k)!}{(k!)^{2}}\,y^{k} = 
\mc O(y^{K+1})$.\footnote{
Alternatively, the Borel  transform of $F_3(y)$ in \rf{c1}   is given by 
\be \la{cb}\notag
F^B_3  (y) \equiv \sum_{k=0}^{\infty}{  b_k\ov k!} \,y^{k} =\te  {}_2F{}_1\big({1\ov3}, {2\ov 3}, 1, 27 y\big) \ ,\ee
and thus the resulting Borel  sum of  $F_3(y)$ is  expressed in terms of   the  Bessel $K$-function  
\be \la{cb2}\notag
\widetilde F^B_3(y) = \int_{0}^{\infty} dt\,e^{-t}\,  F_B(y t)
= \tfrac{1}{3}\,\sqrt{-\tfrac{1}{3\pi}\,y^{-1}}\ e^{-\frac{1}{54}\,y^{-1}}\,
K_{1\ov 6}\big(-\tfrac{1}{54}\,y^{-1}\big) \ . \ee
 $\widetilde F^B_3(y)$ has an imaginary part for $y>0$  which  vanishes
exponentially fast for $y\to 0^{+}$ and does not affect the asymptotic expansion of
$\widetilde F^B_3(y)$   which is the same as      $F_{3}(y)$ in \rf{c1}.
}

It thus follows from \rf{mb2},\rf{c1} that, in particular,   for a 4d scalar  field    in the 3-plet representation 
 \be 
 \la{c6}
 Z^{\reptrip}_{\rmS,4}=
 \prod_{m=1}^{\infty}  \sum_{k=0}^{\infty}\frac{(3\,k)!}{(k!)^{2}}\,   m^k  \big[\zp(x^{m})\big]^{2k}     \ , 
 \qquad \ \ \  \zp(x)=z_{\rmS,4}(x) =  { x (1+x)\ov (1-x)^3} \ , 
 \ee
which gives   a   closed expression for the  all-order  generalization of  the  leading terms in the 
 expansion   of $Z^{\reptrip}_{\rmS,4}$   
 in \rf{3.9}.  

 As  follows  from \rf{mb2}, the same  general expression \rf{c6} 
 applies also to other fields 
with the corresponding one-particle partition  functions $\zp$  (with statistics accounted for by \rf{fff}). 
It thus encodes  the number of singlet operators  built out  of the  field $\Phi$ in 3-plet representation
of   any  integer  dimension. 
Given $Z^{\reptrip}(x)$   and using the inversion formula in Appendix \ref{app:II}  one 
can find also the  corresponding  single-trace partition function in \rf{333}
that counts the number  of  singlet operators  represented by  irreducible contractions. 
 
 Similar   computation    can be   carried  out in the case of a 
 $p$-tensor field  with  $p$  distinguished  indices    transforming under separate $U(N)$   groups:
   the singlet partition function  $Z$  is then defined  by  gauging the  full $[U(N)]^p$  group (see   \ci{Klebanov:2016xxf}). 
  In this case there are less singlet operators  but again the  large $N$ limit of the small $x$ 
  expansion of $Z$    becomes   only asymptotic starting with $p=3$  case  (see Appendix \ref{app:FF}).


\section{Large  $N$  partition function   and phase transitions}
\la{sec4}

\subsection{Overview}
\la{sec:overview}

In the previous section  we  directly computed the $N=\infty$ limit of the 
 small $x$ (small temperature)  expansion of the 
 partition function  and observed the rapid growth 
of the number  of states   with  the increase of  dimension of the  $U(N)$  representation. 
This suggests a  non-trivial  dependence of $Z$  on the temperature  
with possible  phase transitions. 

Let us  recall   what happens in the  well-known case of the adjoint representation. 
If the typical number of states relevant for thermodynamics at  certain $\beta=1/T$ is 
much smaller than $N^{2}$, 
then   
(\ref{3.3}) is a good approximation to the exact partition function.
 This cannot be true at any temperature. 
While the expression  (\ref{3.3}) is well defined at low temperatures 
it diverges as soon as $z_{\Phi}(x)$ becomes of order 1. 
On general grounds, we can prove that the equation 
$z_{\Phi}(x)=1$ has a 
unique solution $x_c= e^{-\beta_{c}}\in (0,1)$.\footnote{
We have always $z_{\Phi}'(x)>0$, $z_\Phi(0)=0$. Besides,  $z_\Phi(1) >1$ if there are at least two single-particle states.
} Then $Z(\beta)$ 
in (\ref{3.3}) is 
well defined for $\beta > \beta_{c}$ and diverges $Z\sim (\beta-\beta_{c})^{-1}$ for 
$\beta\to \beta_{c}$.\footnote{This is consistent with the Hagedorn behaviour $\rho(E) \simeq 
e^{\beta_{c}\, E}$ where $\rho(E)$ is the density of states  appearing in the 
 the partition function   written as 
 $Z=\int  dE\,  \rho(E)\,x^{E} $.  The existence of the maximal  temperature 
$T_{c}=1/\beta_{c}$
is an unphysical   feature  which is a consequence   of  the fact  that 
 the assumption that  the  number of relevant states  is much smaller  than 
  $N^{2}$ fails around $T_{c}$.}
  
  \iffa
In the vector representation  case,  at finite $N$   singlet bilinears  with  many derivatives 
are not all independent and there are also 
corrections to (\ref{3.1}) due to additional singlets that are not bilinears  but involve  factors of the invariant tensor 
$\epsilon_{i_1...i_N}$  \cite{Shenker:2011zf}. 
\fi

{A correlation  between  increasing $N$ and 
the  temperature  
is  clear also from the method   we used to 
 derive the  $N\to \infty$  limit of the small $x$ expansions of $Z$ 
  in section \ref{sec3}.
As we remarked there, the coefficients of these  expansions come from the evaluation of group integrals 
like (\ref{3.4}) and these are independent of $N$ as soon as it  is larger than some  number depending on the degree $\kappa$ in \rf{344}.
As  the temperature is increased so that  $x\to 1$,  more terms in the 
small $x$ expansion are needed 
to accurately describe $Z$. 
However, at higher orders in $x$, the typical $\kappa$  appearing in the 
computation increases  and one needs to go to larger (but finite)  values of   $N$ to  
match   the $N=\infty$ limit. 
This  indicates  that there is a  tension between  increasing 
the temperature and taking the large $N$ limit.}


\def \g  {\gamma}\def \te {\textstyle}

The standard way to  systematically  describe 
 what happens as  the temperature is increased is to   consider  the 
 distribution of eigenvalues of  the group element 
$U$ in (\ref{2.5}) that dominates the partition function 
at  certain temperature.
 At large $N$,   the eigenvalue distribution 
may be approximated by a continuous density  $\rho(\alpha)$ with $\alpha\in(-\pi,\pi)$
obeying
\be
\la{4.1x}
\rho(\alpha)\ge 0, \qquad\qquad \int_{-\pi}^{\pi}d\alpha\,\rho(\alpha)=1.
\ee
In general, 
there may be transition points when one passes from a phase  where 
 $\rho>0$  is  non-zero everywhere on $ (-\pi,\pi)$  to a phase where 
$\rho> 0$ only on an  interval $(-\alpha_{0},-\alpha_{0})\subset (-\pi,\pi)$
(see, {\em e.g.},  \cite{Jurkiewicz:1982iz}).
 This is   what happens in  the vector and adjoint  cases 
\cite{Shenker:2011zf,Aharony:2003sx}.
 On general grounds, a transition   may  be expected 
when in (\ref{2.5}) there is balance between 
the temperature-independent measure  term   $\sim N^{2}$  
and the character-dependent  term in the exponent. 

As we   will   see  below,  this   leads 
to a simple condition 
\be
\la{4.2x}
N^{2} \sim  N^{p}\,z_{\Phi}(x_{c}),\qquad \qquad x_{c} = e^{-1/T_{c}},
\ee
where  $p=1,2,3, ...$ for the vector, adjoint,  3-plet representation, etc.,  cases.
 In the vector case, $x_{c}\to 1$
as $N\to \infty$ and taking into account that $z_{\Phi}(x)\stackrel{x\to 1}{\sim} T^{d-1}$
(cf. \rf{2.2}) 
we find that   \be 
\la{4.3x}
T_{c}^{\repvec}\sim N^{\frac{1}{d-1}} \  \gg 1 \ . \ee
In the adjoint case, (\ref{4.2x}) gives  $T_c$ which is  independent of $N$, 
\be \la{4.4x}
T_{c}^{\repadj}\ \sim 1\ . \ee
In the 3-plet case, we are  then to expect 
$T_{c}$ to vanish as $N\to \infty$ in a way depending on a detailed small $x$ behaviour of 
 $z_{\Phi}(x)$. 
For example, for a  scalar field theory in $d$ dimensions we find from \rf{2.2}
\be
\la{4.5x}
 T_{c}^{\reptrip} \ \sim\  {\te  \frac{d-2}{2}}\,\frac{1}{\log\,N}\  \ll 1 \ .
\ee
The   scalar contribution is  dominant  in the $x\to 0$  limit  also when one considers 
 a collection of fields in \rf{2.2},\rf{202}  like  the   (2,0) tensor multiplet in 6d,
  where thus $T_{c}^{\reptrip} \sim   \frac{10 }{\log\,N}$. 
  Note also that,  as follows from \rf{4.2x},  in the $p>3$ -plet case  we will also get 
  $T_{c}^{p\rm-plet} \sim   \frac{1 }{\log\,N}$.


The physical reason why $T_{c}\to 0$  at large $N$ in the case of the 
 3-plet representation 
  is that the number of states (operators) 
grows too  quickly with increasing energy (dimension).\foot{A possible analogy is  
 with a  1d Ising model which has $T_{c}=0$ when the length  $N$ of the spin 
chain is infinite. There the aligned  "all spins up" (or  "all spins down") 
 configuration $\uparrow\uparrow\dots\uparrow$
  has a finite energy gap  compared to  each of the 
  configurations of the form 
$
\uparrow\uparrow\dots \uparrow\downarrow\downarrow\dots\downarrow$. 
Thus, for large $N$, the variation of the 
 free energy $\Delta F=\Delta {{\cal E}}-T\,\Delta S\sim -T\,\log N$ 
is negative and the entropy term destabilises the ordered configuration for  any $T$.}
Details of this picture  will be  worked out in    the remainder of this   section.
In particular, 
in the  3-plet case  we will  confirm that the phase structure  depends 
on   the  value of  $N\,z_\Phi(x)$ as implied by \rf{4.2x}.

\def \zp {z_\Phi}

At large $N$, we shall find a first order discontinuous transition between a phase 
(where  $N\,\zp$ is smaller than a critical value $(N\, \zp)_{c}$)
with  $\rho>0$ everywhere on $ (-\pi,\pi)$  
 and a phase (where    $N\,\zp>(N\, \zp)_{c}$)
with  $\rho>0$ only for $|\alpha|\le \alpha_{0}$ with $\alpha_{0}\sim (N\,\zp)^{-1/2}$.
We shall  find that  the transition point  corresponds to 
 $(N\, \zp)_{c} = \frac{9}{16}$. 
 At any
temperature,  for a  sufficiently large $N$  the   system will 
 be in  the second phase.
  This means that at  large $N$ the  critical temperature is approaching  zero. 

\subsection{Large $N$ limit in terms of  eigenvalue density}

The integration over $U$  in (\ref{2.5})  may be represented in terms of the eigenvalues of the 
unitary matrix  $\{e^{i\,\alpha_{i}}\}$ with ${-\pi < \alpha_{i} \le \pi}$
\be
\la{4.1}
\int dU = \prod_{i=1}^{N}\int_{-\pi}^{\pi}d\alpha_{i}\,\prod_{i<j}\sin^{2}{\te
\frac{\alpha_{i}-\alpha_{j}}{2}} \ ,\qquad\qquad  \text{tr}(U) = \sum_{i}e^{i\,\alpha_{i}}.
\ee
Using the explicit  form of the characters in (\ref{2.6}), we  get 
\begin{align}
&\la{4.2}\qquad \qquad 
Z  
= \int d\bm{\alpha}\  e^{-S(\bm{\alpha}, x)}\ ,
\\
&
\la{4.3}
S({\bm\alpha}, x) = -\frac{1}{2}\,\sum_{i\neq j}\log\sin^{2}{\te \frac{\alpha_{i}-\alpha_{j}}{2}}
+\sum_{m=1}^{\infty} c_{m}(x) \,\mc V(m\,\bm\alpha), \qquad \qquad c_{m} \equiv -\frac{1}{m}\,z_\Phi (x^{m}),\end{align}
where 
\begin{align}
{\mc V}^{\repvec}(\bm\alpha) &= 2\,\sum_{i=1}^{N} \cos \alpha_{i}\ , \la{4.9} \\
{\mc V}^{\repadj}(\bm\alpha) &= \sum_{i,j=1}^{N} \cos(\alpha_{i}-\alpha_{j})\ , \la{4.10}\\
{\mc V}^{\reptrip}(\bm\alpha) &= 2\,\sum_{i,j,k=1}^{N} \cos(\alpha_{i}+\alpha_{j}+\alpha_{k})\ . \la{4.100} 
\end{align}
The extremum condition for  $S({\bm\alpha}, x) $  in   (\ref{4.3}) is, {\em e.g.},  in  the 3-plet case
\be
\la{4.7}
\sum_{j\neq i}\cot{\te \frac{\alpha_{i}-\alpha_{j}}{2}}-6\,\sum^\infty_{m=1}
z_\Phi (x^{m})\,\sum_{j,k=1}^{N} \sin\big[m(\alpha_{i}+\alpha_{j}+\alpha_{k})\big]=0.
\ee
Let us follow  \cite{Aharony:2003sx,Shenker:2011zf,Giombi:2014yra} 
and replace the integration over 
$\bm\alpha$ by the integration over the eigenvalue  density $\rho(\alpha)$ which is a periodic 
function 
on the unit circle $\alpha\in(-\pi,\pi)$\footnote{Here 
the Dirac delta function is the periodic one: \ \ 
 $\delta(\alpha)\to \sum_{k\in\mathbb Z}\delta(\alpha+2\pi k)$.}
\be
\la{4.8}
\rho(\alpha) = \frac{1}{N}\,\sum_{n=1}^{N}\delta(\alpha-\alpha_{i}) \ . 
\ee
It satisfies the  conditions  (\ref{4.1x})  by construction.
Then the  action  in \rf{4.3}  becomes
 \be
S(\rho,x) = S_{M}(\rho)+V(\rho,x) \ ,\la{4.14}
\ee
with  the measure term 
\begin{align}
&S_{M} = N^{2}\,\int d\alpha\,d\alpha'\,K(\alpha-\alpha')\,\rho(\alpha)\,\rho(\alpha'),\la{4.15} \\
&
K(\alpha) = -\frac{1}{2}\,\log(2-2\,\cos\alpha) = \sum_{m=1}^{\infty}\frac{1}{m}\cos(m\alpha) \ , 
\la{4.166}\end{align}
and the potential term 
\begin{align}
\la{4.11}
V^{\repvec} &= 2\,N\,\int d\alpha\, \rho(\alpha)\,\sum^\infty_{m=1}c_{m}(x)\,\cos(m\,\alpha),  \\
V^{\repadj} &= N^{2}\,\int d\alpha\,d\alpha' \rho(\alpha)\,\rho(\alpha')\,\sum^\infty_{m=1}c_{m}(x)\,
\cos\big[m\,(\alpha-\alpha')\big], \la{4.111}\\
V^{\reptrip}&= 2\,N^{3}\,\int d\alpha\,d\alpha'\,d\alpha''\, \rho(\alpha)\,\rho(\alpha')\,\rho(\alpha'')\,\sum^\infty_{m=1}c_{m}(x)\,\cos\big[m\,(\alpha+\alpha'+\alpha'')\big].\la{4.19}
\end{align}
 The integral  expression for $Z$ in \rf{4.2}   can then  be  written as a 
 path integral over the field $\rho(\alpha)$  subject to the constraints 
 \rf{4.1x}   with the action \rf{4.14}.

As the constant part of $\rho(\alpha)$  drops  out of \rf{4.15}--\rf{4.19}, the   
 $\rho(\alpha)=\frac{1}{2\pi}$
 is always a  stationary point of the action \rf{4.14}.
Let us  first  briefly review the vector and adjoint cases  where 
 a perturbative expansion   around this  constant density is sensible
 and then  turn to  the 3-plet case where  this is not the case. 

\subsection{Vector and adjoint cases}

One may  expand $\rho(\alpha)$ in Fourier modes as 
\be
\la{4.12}
\rho(\alpha) = \frac{1}{2\,\pi}+\frac{1}{N }\Big[ {1\ov \pi} \,\sum_{m=1}^{\infty}\rho_{m}^{+}\cos(m\alpha)
+\frac{1}{\pi}\,\sum_{m=1}^{\infty}\rho_{m}^{-}\sin(m\alpha)\Big].
\ee
This expansion is meaningful if the $\{\rho_{m}^{\pm}\}$ variables turn out to
have  a  some stationary point with
 bounded fluctuations. Then, at large $N$,  the positivity
constraint on $\rho$ in (\ref{4.1x}) will not  be violated. 
Written in  terms of the variables $\{\rho_{m}\}$  the measure term \rf{4.15} becomes 
\be\la{4.211}
S_M=  
  \,\sum_{m=1}^{\infty}
\frac{1}{m}\,\big[(\rho_{m}^{+})^{2}+(\rho_{m}^{-})^{2}\big]\ , 
\ee
while  the potentials in \rf{4.11} and \rf{4.111} take the form 
\be\la{4.222}
V^{\repvec} = 2\,\sum_{m=1}^{\infty} c_{m}\,\rho_{m}^{+}\ , \qquad\qquad 
V^{\repadj} = \,\sum_{m=1}^{\infty} c_{m}\,\big[(\rho_{m}^{+})^{2}
+(\rho_{m}^{-})^{2}\big].
\ee
In the {vector} case, the resulting  action \rf{4.14}  is stationary at  
\be
\la{4.155}
\rho_{m}^{+} = -\,m\,c_{m}=z_\Phi (x^{m})  \ ,\qquad \qquad \rho_{m}^{-}=0\ ,
\ee
and, as a result, we find the same expression for $Z$  as given earlier in \rf{3.1} 
\be
\la{4.16}
\log Z^{\repvec} =
\sum_{m=1}^{\infty} \frac{1}{m}\,
\big[z_\Phi (x^{m})\big]^{2}\ .
\ee
The same expression is found  by doing the Gaussian integral over $\rho_{m}^{\pm} $ 
 in \rf{4.2}. 
 
In the {adjoint} case the total action is 
\be
\la{4.17}
S^{\repadj} = \,\sum_{m=1}^{\infty}\frac{1+m\,c_{m}}{m}\,\big[(\rho_{m}^{+})^{2}+
(\rho_{m}^{-})^{2}\big]=    \,\sum_{m=1}^{\infty}\frac{1-z_\Phi (x^m) }{m}\,\big[(\rho_{m}^{+})^{2}+
(\rho_{m}^{-})^{2}\big] \ . 
\ee
Integrating  over $\rho_{m}^{\pm} $ 
 in \rf{4.2}  with the  normalization $Z(x=0)=1$, we then obtain 
\be
\la{4.18}
Z^{\repadj} = 
\prod_{m=1}^{\infty}\big[1-z_\Phi (x^{m})\big]^{-1}\  \ \ \to\ \ \ 
\log Z^{\repadj} = -\sum_{m=1}^{\infty}\log\big[1-z_\Phi(x^{m})\big],
\ee
which is the same as   (\ref{3.3})   quoted earlier.  

 A  similar   derivation of $Z$   applies 
in the   case   when the adjoint representation is 
replaced by 2-plet  one  $N\otimes N$   or   its  (anti) symmetric  version 
(see Appendix \ref{app:A}).

\medskip
As  discussed in   section \ref{sec:overview}, 
if  the temperature  (and thus $x$ and $z_\Phi(x)$) are 
 small enough, the expressions in (\ref{4.16}) and (\ref{4.18}) converge and 
 the expansion (\ref{4.12}) is well behaved and describes a strictly positive density.
Increasing the temperature we reach a 
transition point 
where $\rho$ starts to be zero only on a subset $(-\alpha_{0},\alpha_{0})\subset(-\pi,\pi)$.
In the vector case, this transition happens at $T_c \sim N^{1\ov d-1}$ as in \rf{4.3x}. The expansion (\ref{4.12}) is no longer  valid  above this critical temperature.
 In adjoint case, 
the critical temperature is determined by the 
 condition $z_\Phi (x)=1$  when the 
fluctuations of modes in \rf{4.17} are no longer suppressed.  Thus in this case  we get 
the critical temperature $T_{c}\sim 1$ in  \rf{4.4x}  which is   independent of $N$. 
Details of the  form of 
$\rho(\alpha)$ in the two phases for the  vector and adjoint 
models  may be found in    \cite{Shenker:2011zf,Aharony:2003sx,Amado:2016pgy}.

In the low temperature   phase where   the expansion \rf{4.12} is  valid, the  value of the action \rf{4.14} 
({\em i.e.} the measure term  \rf{4.211}  plus  potential in \rf{4.222})
  and thus  the resulting large $N$  partition function \rf{4.16}  or \rf{4.18} 
 does not depend on $N$, {\em i.e.} $\log Z = \mc O(N^0)$.

The stationary  point of the action \rf{4.14}  is  given by  the  analog  of eq. \rf{4.3}     written  in terms of $\rho(\a)$ (cf.  \rf{5.2}  below).  Assuming it admits  a  non-trivial  solution for $\rho(\a)$ 
(due to a balance between the measure and potential contributions   to the stationary point   equation)
then  in the  higher temperature $T >T_c$ phase  the resulting  value of the action \rf{4.14} 
at the stationary point will scale   as the   measure term, {\em i.e.}  as $N^2$.
 Thus  the  stationary-point approximation will 
 be valid in the large $N$ limit   with $\log Z = \mc O(N^2)$  in the high-temperature phase  (see  \rf{D.5},\rf{D.55}). 
 
In the 3-plet case   discussed below the  low temperature  phase  will  be shrinking with increasing $N$  (with $T_c \to 0$)
while the  action and  $\log Z$   will be scaling again as $N^2$    at the non-trivial  stationary-point solution of  \rf{5.2}.

\subsection{3-plet case}

In the 3-plet case, using \rf{4.12} in the potential term \rf{4.19} we get 
\be
\la{5.1}
V^{\reptrip} = 2\sum_{m=1}^{\infty} c_{m}\,\big[
(\rho_{m}^{+})^{3}-3\,\rho_{m}^{+}\,(\rho_{m}^{-})^{2}
\big]\ . 
\ee
The resulting action given by the sum of \rf{4.211}   and  (\ref{5.1}) is {unbounded}
 from below (in particular,  the analog of \rf{4.2}  giving  $Z$ as an 
  integral over $\rho^\pm_m$  does not converge). 
  Thus in the 3-plet  case the  constant  density $\rho=\frac{1}{2\pi}$ is really a saddle and  not a minimum even at low temperatures;
the Fourier coefficients $\rho_{m}^{\pm}$  in \rf{4.12} 
tend to be large  and this violates
 the fundamental positivity condition on 
$\rho(\alpha)$ in \rf{4.1x}. 

A phase transition from the trivial  homogeneous phase $\rho=\frac{1}{2\pi}$ 
should happen when the 
cubic in $\rho$  term  in the action  \rf{4.14},\rf{4.19} 
 becomes  of order   of  the quadratic  measure one in \rf{4.15}. 
In the vector case the transition point  is when 
we   balance a measure term \rf{4.15} 
 $\sim N^2$ against the  potential term \rf{4.11}  $\sim N z_\Phi(x)$. In the adjoint case,
both  terms \rf{4.15} and \rf{4.111}  are  of the same order 
$\sim N^{2}$. In the 3-plet case 
the potential term \rf{4.19}  scales as  $N^{3}z_\Phi(x)$ and  we are led to the condition 
in \rf{4.2x}
(in a sense,  the  vector and the 3-plet cases are opposite  to each other  with 
the  adjoint case  being  in between). 

Below we will   explore  the 3-plet case  in more detail, by first   studying 
 numerically the eigenvalue 
distribution and  then  presenting some  analytical expressions.

%
\def \a  {\alpha}

\subsubsection{Numerical analysis}

\begin{figure}[t]  
  \centering
 \includegraphics[scale=0.35]{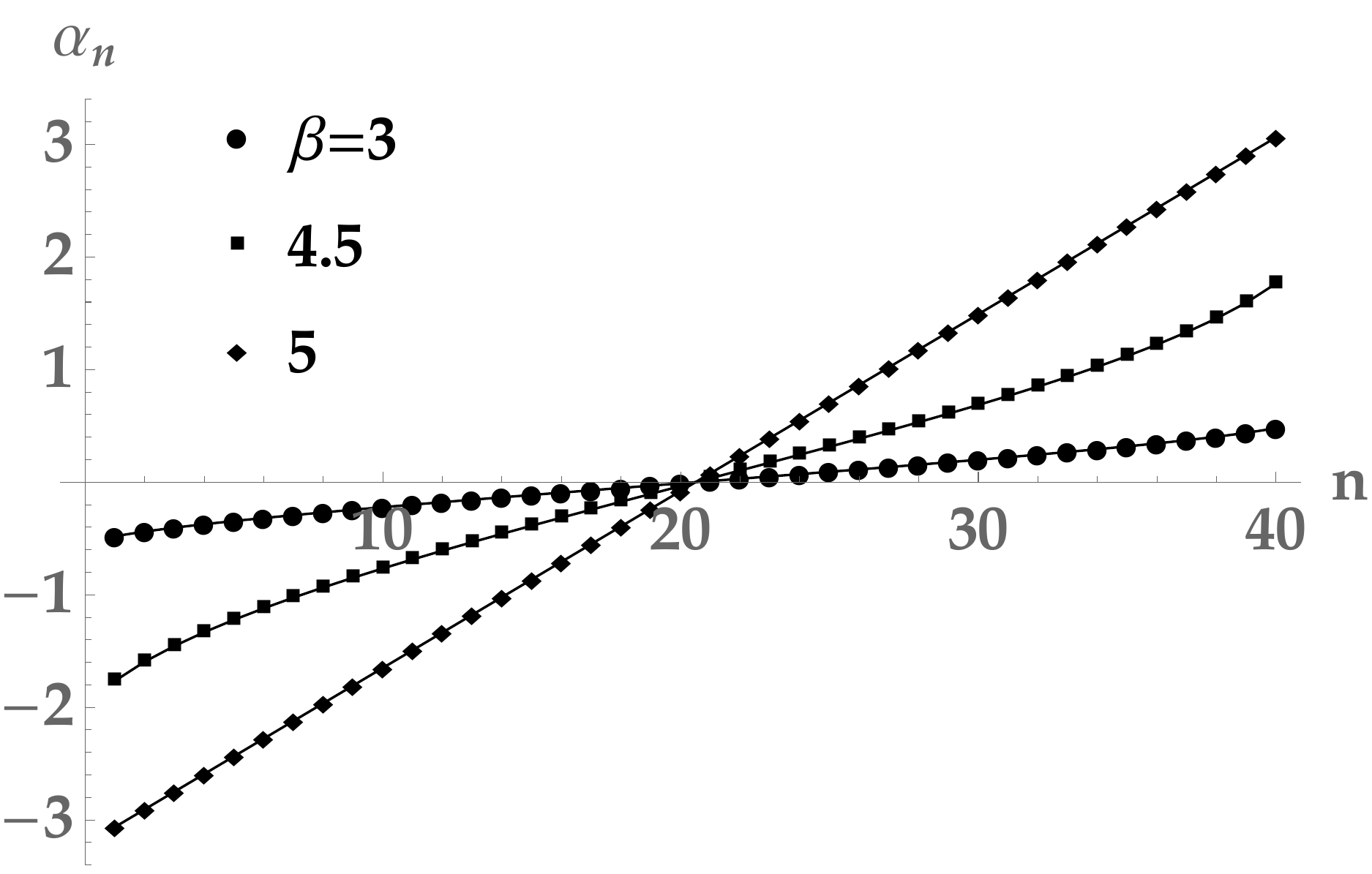}
\includegraphics[scale=0.35]{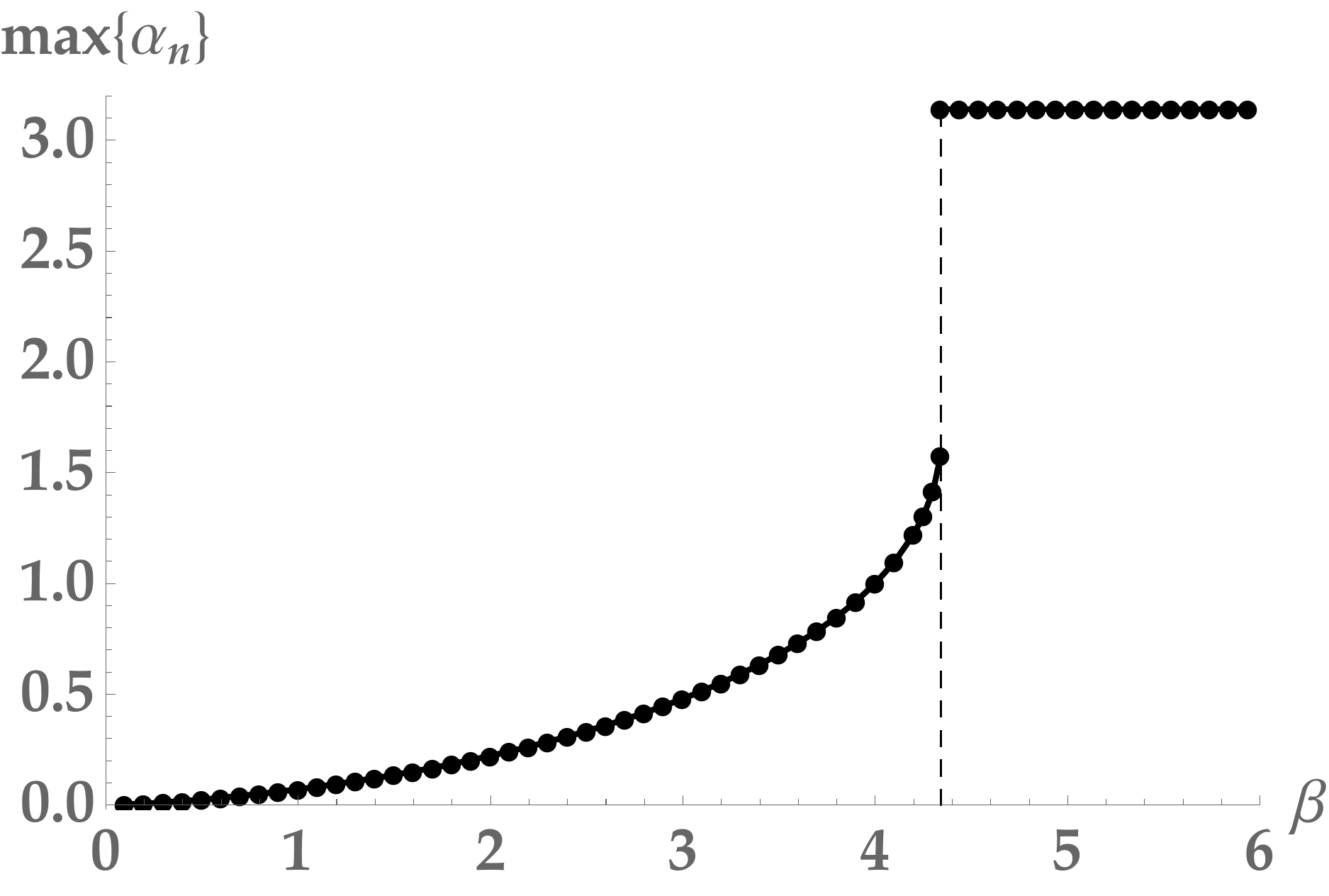}
   \caption[]
   {
   Exact eigenvalues for $N=40$, $M=4$ and three values of $\beta$ well below, just   below and just 
   above  the transition.  The right graph represents 
   the   plot of  the maximal  eigenvalue that  demonstrates  that it jumps at a certain 
   temperature from a value $\simeq{ \pi\ov 2}$ to $\pi$.
    \label{fig1}  }
\end{figure}

As a preliminary step, 
we can try to determine numerically the solution to the extremum condition (\ref{4.7}). 
We shall sum over $m$ in \rf{4.7} with $1\le m \le M$ to   investigate    the effect of 
higher harmonics. We shall consider  the case of 4d scalar with $\zp(x)= {x(1+x)\ov (1-x)^3}$  (see \rf{2.2}). 
Solving numerically (\ref{4.7}) with $N=40$, $M=4$ we find the following behaviour:
\begin{enumerate}
\item At high temperature, the eigenvalues $\a_i$ 
are distributed on   $(-\alpha_{0}, \alpha_{0})\subset (-\pi,\pi)$
with $\alpha_0 \to 0$ as $\beta\to 0$. Their density is approximately 
 flat  near  the center of the interval, but with deviations at the 
edges.
\item As the temperature is reduced,  there is a  transition 
point where the distribution becomes  essentially uniform over the whole interval $(-\pi, \pi)$. 
\end{enumerate}
This is illustrated  in Fig.~\ref{fig1} where we show in the left graph 
the exact eigenvalues at $\beta=3, 4.5, 5$ that are
well below, just below and  above the  transition. The right graph 
shows $\max |\alpha_{n}|$ as a function of $\beta$ to emphasize the rapid jump  
from a value $\simeq {\pi\ov 2}$ to $\pi$ at a critical temperature. In
Appendix  \ref{app:L} we  will consider explicitly the   small  $N=2$   case 
  where these features are already visible. 


\subsubsection{Exact solution for the eigenvalue density at large $N$}

\begin{figure}[t]  
  \centering
 \includegraphics[scale=0.5]{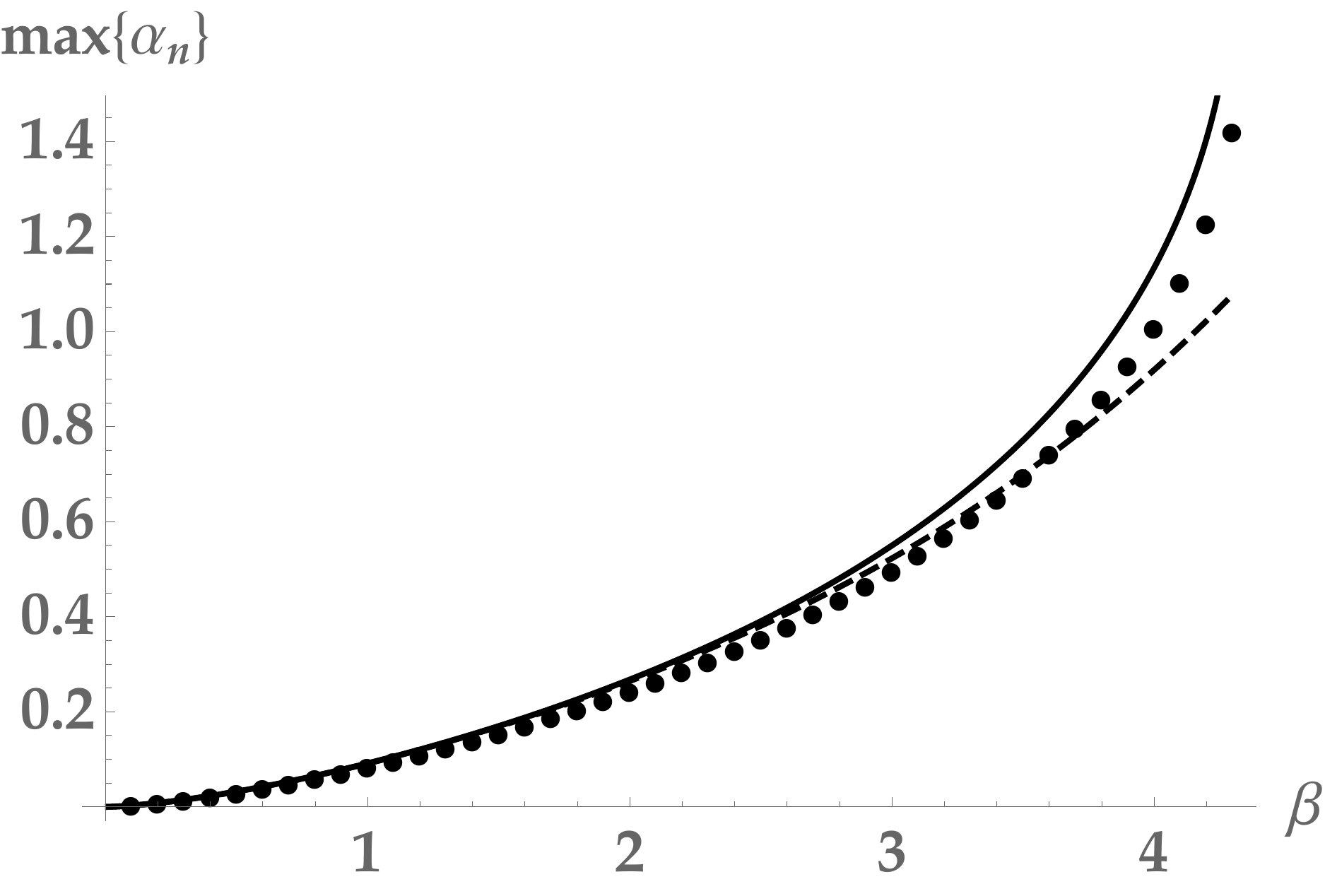}
   \caption[]
   {
   Comparison between the 
edge of the exact eigenvalue distribution at $N=40$ in  the case of 
one harmonic in the potential term  in \rf{4.7} (solid circles) with  the solution of (\ref{5.17}) (solid line). We also show the high 
temperature approximation (\ref{5.18}) in the one-harmonic  case 
(dashed line). 
    \label{fig5}  }
\end{figure}

The features observed in the numerical analysis may be explained  analytically 
 as follows. Let us return to the stationary point condition in 
eq. (\ref{4.7}) expressing  it in terms of $\rho(\a)$: 
\be\!\!\!\!
\la{5.2}
\int d\alpha' \,\rho(\alpha')\,\cot {\te \frac{\alpha-\alpha'}{2}}
=6N\sum^\infty_{m=1}\,\zp(x^m)
\int d\alpha' d\alpha''\,\rho(\alpha')\,\rho(\alpha'')\,\sin\big[m\,(\alpha+\alpha'+\alpha'')\big].
\ee
Let us assume that $\rho$ is symmetric and 
supported on $(-\alpha_{0},\alpha_{0})$  and thus  write (\ref{5.2}) as 
\begin{align}
\la{5.3}
\int d\alpha' \,\rho(\alpha')\,\cot{\te \frac{\alpha-\alpha'}{2}}
 &= 6\,N\,\sum^\infty_{m=1}\,z_\Phi(x^m)\,
\int d\alpha'\,d\alpha''\,\rho(\alpha')\,\rho(\alpha'')\,\sin(m\,\alpha)\,\cos\big[m\,(\alpha'+\alpha'')\big]\notag \\
&= 2\,\sum^\infty_{m=1} a_{m}\,\rho_{m}^{2}\,\sin(m\,\alpha),
\end{align}
where
\be
\la{5.4}
a_{m} = 3\,N\,z_\Phi(x^m),\qquad\qquad  \rho_{m} = \int d\alpha \,\rho(\alpha)\,\cos(m\,\alpha).
\ee
Eq. \rf{5.3}  is same as eq.~(5.20) of \cite{Aharony:2003sx}  and thus  its solution may be written as
\begin{align}
\la{5.5}
\rho(\alpha) &={\te  \frac{1}{\pi}\,\sqrt{\sin^{2}\frac{\alpha_{0}}{2}-\sin^{2}\frac{\alpha}{2}}\ }
\sum_{k=1}^{\infty} Q_{k}\,\cos\big[(k-\tfrac{1}{2})\,\alpha\big], \\
\la{5.6}
Q_{k} &= 2\,\sum_{\ell=0}^{\infty} a_{k+\ell}\,\rho^{2}_{k+\ell}\,P_{\ell}(\cos\alpha_{0}),
\end{align}
where $P_\ell$ is the Legendre polynomial. 
To simplify the presentation, let us  first consider a model   with   just  one harmonic $\rho_1$ 
 present   in the r.h.s. of (\ref{5.3}) which   should 
be  a good approximation for large $\beta$  when $x= e^{-\b} \ll 1$   and thus the value of $a_m$ in \rf{5.4}   decreases with $m$.
 Then  
\be
\la{5.7}
\rho(\alpha) = \te \frac{2}{\pi}\, a_{1}\,\rho_{1}^{2}\,\sqrt{\sin^{2}\frac{\alpha_{0}}{2}-\sin^{2}\frac{\alpha}{2}}\ 
\cos\frac{\alpha}{2}.
\ee
We still  need to impose  the  self-consistency equations  
\be
\int_{-\alpha_{0}}^{\alpha_{0}}d\alpha\,\rho(\alpha) = 1, \qquad\qquad 
\int_{-\alpha_{0}}^{\alpha_{0}}d\alpha\,\rho(\alpha)\,\cos\alpha = \rho_{1}.
\ee
We then find 
\begin{align}
& 2\,a_{1}\,\rho_{1}^{2}\,u = 1, \qquad \qquad\qquad   \la{5.10}
 2\,a_{1}\,\rho_{1}^{2}\,u\,(1-\tfrac{1}{2}\,u) = \rho_{1} \ , \qquad  u\equiv \sin^{2}\tfrac{\alpha_{0}}{2}\in(0,1)\\
&\la{5.11}
1-\tfrac{1}{2}u = 
(2\,a_{1}\,u)^{-1/2} \ , \qquad \qquad a_1= 3N\, \zp (e^{-\b}) \ .
\end{align}
Thus $u$ solves  a cubic equation.
For large $a_{1}$ we get a consistent  solution 
\be
\la{5.12}
u=\frac{1}{2\,a_{1}}+\frac{1}{4\,a_{1}^{2}}+\dots \quad \rightarrow \quad
\alpha_{0} = \big[\tfrac{3}{2}N\,z_\Phi(e^{-\beta})\big]^{-1/2}+\dots.
\ee
As $a_{1}$ decreases, we find a solution with $u\in (0,1)$ only up to the point
\be
\la{5.13}
a_{1}=\frac{27}{16}\ ,   \qquad  {\rm i.e.}  \qquad u=\frac{2}{3}\ . 
\ee
This limiting value corresponds to the  maximal  width of the interval  being 
\be
\alpha_{0} = 2\,\arctan\sqrt{2}\  \simeq \ 0.6\,\pi.
\ee
{To summarize, } including just  one harmonic in the sum in  \rf{5.3}, 
for each temperature and $N$ such that 
\be
N\,z_\Phi(e^{-\beta})   > \frac{9}{16} \ , 
\ee
we get  the eigenvalue distribution 
\be
\rho(\alpha) = \begin{cases}
\frac{1}{\pi\,\sin^{2}\frac{\alpha_{0}}{2}}\,\sqrt{u-\sin^{2}\frac{\alpha}{2}}\,
\cos\frac{\alpha}{2} \ , & \ \ \ |\alpha|<\alpha_{0} \\
0\   &\ \ \    \alpha_{0} < |\alpha|\leq \pi 
\end{cases}
\ee
where $u=\sin^{2}\frac{\alpha_{0}}{2}$ is determined by the relation (\ref{5.11}), \ie
\be
\la{5.17}
\tfrac{3}{2}\,u\,(2-u)^{2} = \big[N\,z_\Phi(e^{-\beta})\big]^{-1}.
\ee 
For $\frac{9}{16}<N\,\zp(e^{-\b})<\frac{2}{3}$ there are two solutions $u_1,u_2$ with 
 $0<u_{1}<u_{2}<1$.
Here $u_{1}$ is a minimum of the action, while $u_{2}$ is a local maximum, see 
Appendix \ref{app:K}. 
A numerical test of (\ref{5.17}) is shown in Fig. \ref{fig5} where we compare its prediction with  the 
edge of the exact eigenvalue distribution at $N=40$ 
 found   by  taking  just one harmonic in the potential term in \rf{4.3}.

Including up to $M$ higher harmonics, the equations (\ref{5.10}) are replaced by similar ones involving 
$u$ and $\rho_{1}, \dots, \rho_{M}$. The solution depends on $N$ and $\beta$  through the 
combinations 
  $a_m=3N\,z_\Phi(e^{-m\,\beta})$, $\ m=1, \dots, M$ (see \rf{5.3},\rf{5.4}). For large $N$, the critical inverse
temperature $\beta_{c}(N, M)$ admits a finite limit that may,  in principle,  
 depend on $M$.  In fact, $a_{c}$   defined as 
\be
\la{mbx1}
a_{c}(M) \equiv  \lim_{N\to \infty} 3N\,z_\Phi(e^{-\beta_{c}(N, M)})\ .
\ee
 is independent of
 $M$. To see why, consider,  for example,   $\zp(x)= { x(1+x)\ov (1-x)^3}$ corresponding to 
   4d scalar theory.
 If we solve $a_{c} = 3N\,z_\Phi(e^{-\beta_{c}})$ for $\b_c$ at large $N$, we get
$\beta_{c} = -\log{a_{c}\ov 3N}+{4\,a_{c}\ov 3N}-{5\,a_{c}^{2}\ov 3N^{2}}+\dots$. The coupling
parameters $a_m=3N\,  \zp(e^{-m\b})$   for  higher 
harmonics in \rf{5.3},\rf{5.4} are then subleading, e.g.,  $3N\,z_\Phi(e^{-2\,\beta_{c}}) =
{ a_{c}^{2}\ov 3N}-{8\,a_{c}^{3}\ov 9N^{2}}+
\dots$, and in general $3N\,z_\Phi(e^{-m\,\beta_{c}})  = {a_{c}^{m}\ov (3N)^{m-1}}+\dots$. Thus,
\be
\la{mbx2}
a_{c} = a_{c}(M=1) = \frac{27}{16}\ , 
\ee
which is the  value of $a_1$   found in \rf{5.13}.

To give an example, with $M=3$ harmonics and taking $N=50, 60, \dots, 100$ we get 
\be
\la{mbx3}
\begin{array}{ccccccc}
\toprule
N & 50 & 60 & 70 & 80 & 90 & 100 \\ 
\midrule
N\,z_\Phi(e^{\beta_{c}(N, M=3)})\ \ \ \ \  & 0.562392  & 0.562409 & 0.562421 & 0.56243 & 0.562438
& 0.562444\\
\bottomrule
\end{array}
\ee
A quadratic fit with $1/N$ and $1/N^{2}$ corrections gives $a_{c}(M=3) = 3\times 0.562502$
which is equal to $27\ov 16$ in \rf{mbx2}  with a $10^{-6}$ relative precision.

As we will  show in Appendix \ref{app:Q},   deep into the high temperature phase, 
  the 1-harmonic result (\ref{5.12}) is replaced by 
\be
\la{5.18}
\alpha_{0} = \Big[\tfrac{3}{2}\,N\,\sum_{m=1}^{\infty} m\,z_\Phi(e^{-m\,\beta})\Big]^{-1/2}+\dots
\ee
Similar  results  are found  in the case when the  general  3-plet representation is replaced by the 
symmetric or antisymmetric one:  as we  will show in 
Appendix \ref{app:F}, the large $N$ behaviour   is the same  in  all of these  cases. 


\def \OO {{\cal O}}

\section{Concluding remarks}
\la{sec5}


In this paper we   discussed  singlet  partition function  $Z$  of  conformal theories   defined
by free   fields  in higher representations of an internal symmetry group. 
We observed that   starting with rank 3  tensor case  the number of  singlet states     grows so fast  with the energy 
that the small temperature expansion  of $Z$  has   zero radius of convergence in the $N=\infty$  limit. 
This is reflected in the  vanishing of  the  critical temperature $T_c$  at $N=\infty$. 
For  large but  finite  $N$  there are two phases:  $T <T_c$ and $T >T_c$, with 
$\log Z \sim N^2$   in the higher temperature  phase (same  scaling as found in 
 the vector   and adjoint representation cases).

We   have   concentrated on  the   case  of the  $p$-fundamental representation of $U(N)$ 
but  similar    conclusions are    true also  for  the $[U(N)]^p$ invariant  singlet partition function  
 of $p$-tensors  with inequivalent indices  (see Appendix \ref{app:FF}). 
 The same qualitative  behaviour is found  also  when $U(N)$ symmetry is replaced by  $O(N)$   (cf. Appendix \ref{app:X}).

 One open  question  is  about possible implications for the  AdS dual of the  free $p$-plet  or $p$-tensor CFT. 
 The   rich   spectrum   of singlet operators  implies the presence of infinite 
 sequences of massive fields  in AdS (suggestive of a   "tensionless membrane"  spectrum  in the  $p=3$   case). 
 It would   be interesting to  shed further light on this by studying simplest   correlation  functions 
 of  operators   like $ \overline \phi_{ijk}\phi_{ijk}$ (dual to a scalar in AdS)  by generalizing  
 the discussion of  the vector  and adjoint cases  in    \ci{Amado:2016pgy}. 
 
 For the fields in  the vector  and adjoint  representations 
  the  large $N$ 
  free energy or  1-loop $\log Z$ of all  higher spin fields  in thermal AdS    scales as   $\mc O(N^0)$ 
  and   matches the  corresponding boundary  CFT   free  energy   in the   low-temperature phase    in  \rf{3.1}, \rf{3.3}.
  In the high temperature  phase   the boundary CFT free energy    is  $\mc O(N^2)$; 
   that      formally agrees   with  an  AdS black-hole  scaling   in the adjoint   case \ci{Witten:1998qj,Witten:1998zw}. 
   In the vectorial case  where $T_c$ grows  with $N$   and thus the high temperature phase 
   is not obviously attainable, a    possibility of similar  matching 
   remains an open question   \ci{Shenker:2011zf}   (the classical AdS 
    action  here scales as $N$ and thus a classical    thermal object   would contribute
    $\mc O(N)$ to the  free energy). 
    
    In the 3-plet case  the   1-loop partition function in  thermal  AdS  computed  for the  full spectrum of   fields  dual to 
     singlet  conformal operators  in the  large $N$   limit   should also be  expected to be given by an asymptotic series 
     matching the   low temperature phase  expression   for $\log Z (x) = \mc O(N^0)$  in    \rf{c6}.
     The  high temperature phase result $\log Z (x) = \mc O(N^2)$  here appears to be  subleading  to any 
       potential   contribution  coming  from a  classical  AdS action   as  that should scale as $\mc O(N^3)$ 
       (the coefficient in front of the AdS   action  should be $N^3$   to match 
       the   correlation functions in the  free   3-plet   CFT  \ci{Bastianelli:1999ab}).

 Another  important   question is how these  conclusions  may   change in an interacting CFT,  \eg, whether  $T_c$ may become 
finite at a non-trivial  large $N$  fixed point. This   is of particular interest in the  case of the 
 $(2,0)$ tensor multiplet  theory  in 6d that   
 should have an  AdS$_7 $  dual  with a  supergravity limit  in  the $N \to \infty$ limit 
 admitting black holes  and thus predicting    $N^3$ scaling of the free energy    \ci{kt1,Gubser:1998nz}. 


\iffa 
Indeed,   while  the critical  temperatures  in the vector ($T_c \sim N^\alpha , \alpha >0)$ and adjoint ($T_c\sim 1$)
representation  cases are    finite,     
we find that in the 3-plet case  $T_c \sim (\log N)^{-1}$, i.e. it   approaches zero at large $N$. 
We discuss some  details of large $N$ solution for  the eigenvalue distribution. Similar  
 conclusions apply to higher $p$-plet  representations  of $U(N)$     and also to the  free $p$-tensor   theories 
invariant  under $[U(N)]^p$  with $p\geq 3$.  
Instead of planar decomposition -- triangulation of surface --  same for volume -- 4-vertex
tetrahedra -- natural interactions 
\fi 

\def \tr {{\rm tr}}

\section*{Acknowledgments}
We are grateful to  
  I. Klebanov and G. Tarnopolsky 
for very  useful  discussions  and also thank  E. Joung for comments. 
The work of AAT  was   supported by the ERC Advanced grant no. 290456,
 the  STFC Consolidated grant ST/L00044X/1
  and   the Russian Science Foundation grant 14-42-00047 at Lebedev Institute.

\appendix

\section{ $N=\infty$ partition function for  2-plet  representation of $U(N)$}
\la{app:A}

In addition to  the adjoint representation $N\otimes \overline N$ one may consider also  
 another rank 2 tensor representation -- $N\otimes N$  or 
   2-plet  of $U(N)$.
 The corresponding real representation  in 
\rf{2.5} is   $R=N\otimes N+\overline N\otimes \overline N$
and thus  $\chi_R (U) = [\tr\,( U)]^2  + [\tr\,( U^{-1})]^2$ (see \rf{266}).
   The resulting 
potential  in  (\ref{4.3})  is  then  (cf.     \rf{4.9}--\rf{4.100}) 
\be\la{b1} 
V^{\text{2-plet}}(\rho) =  2\,N^{2}\,\int d\alpha\,d\alpha' \rho(\alpha)\,\rho(\alpha')\,\sum^\infty_{m=1}c_{m}\,\cos\big(m\,(\alpha+\alpha')\big).
\ee
In terms of the Fourier coefficients $\rho_{m}^{\pm}$ in \rf{4.12}  we get 
\be
V^{\text{2-plet}}= 2\sum_{m=1}^{\infty} c_{m}\,[(\rho_{m}^{+})^{2}-(\rho_{m}^{-})^{2}].
\ee
As in the adjoint case (\ref{4.17}),(\ref{4.18}),    
integrating over $\rho_m^\pm$    we get for  the   $N=\infty$ partition function 
\be\la{B.3}
\log Z^{\text{2-plet}} 
=-\frac{1}{2}\,\sum_{m=1}^{\infty}\log\big(1-\big[\,2\, z_\Phi(x^{m})\big]^{2}\big) \  ,
\ee
in agreement with \rf{c5}.
The corresponding single-trace partition function is   given in \rf{E.8}. 
The expression \rf{B.3}  is valid in the low temperature  phase, i.e. 
 for temperatures  below the critical one 
where  $ 2\,  z_\Phi(x_c)\sim 1$. 

For  the symmetric 2-plet  representation where 
 $R
 =(N\otimes N)_{\rm sym}+(\overline N\otimes \overline N)_{\rm sym} $    and 
\be
\la{B.4}
\chi_{R}(U) = \frac{1}{2}\big[\,\text{tr}(U)\big]^{2}+\frac{1}{2}\,\text{tr}(U^{2}), 
\ee
we get  instead of \rf{b1} 
\begin{align}
\la{B.5}
V^{\text{2-plet}^+}  &= N^{2}\int d\alpha\,d\alpha' \rho(\alpha)\,\rho(\alpha')\,\sum^\infty_{m=1}c_{m}\,\cos(m\,(\alpha+\alpha'))
+N\int d\alpha\,\rho(\alpha)\,\sum^\infty_{m=1} c_{m}\,\cos(2m\,\alpha) \notag \\
&= \,\sum_{m=1}^{\infty} c_{m}\,\big[(\rho_{m}^{+})^{2}-(\rho_{m}^{-})^{2}\big]
+\,\sum_{m=1}^{\infty}c_{m}\,\rho_{2m}^{+}.
\end{align}
Adding this to  \rf{4.211}   and  performing  again the Gaussian integration over 
 $\rho^{\pm}$ 
 gives (cf. \rf{4.18},\rf{B.3})
\begin{align}
\la{B.7}
\log Z^{\text{2-plet}^+} 
= -\frac{1}{2}\,\sum_{m=1}^{\infty}\log\big(1-\big[z_\Phi(x^{m})\big]^{2}\big)+
\frac{1}{2} \sum^\infty _{m=1}\frac{1}{m}\,\frac{\big[z_\Phi(x^{m})\big]^{2}}{1-z_\Phi(x^{2m})}\ . 
\end{align}
In the antisymmetric 2-plet representation case 
 there is a relative  minus sign in  (\ref{B.4})         and thus  in the last term in \rf{B.5} 
  but  the final 
result  for $Z$  is again the same as  (\ref{B.7}). 

\section{Finite $N$ low temperature expansion  of   3-plet  partition function 
}
\la{app:finiteN}

Here we   will supplement   the large $N$ analysis  in section \ref{sec3}   with a discussion of the  finite  $N$ case. 
At finite $N$, the low temperature expansion of the partition function may  still be done by direct 
expanding (\ref{2.5}). However, simple expressions  like (\ref{3.5}) or 
factorization leading to (\ref{mb1})
are no longer  valid. Instead, the group integrals (\ref{3.4}) must be computed on a case by case basis. 

In particular, for a 4d scalar in 3-plet representation, one can compute the following first five terms 
 in the small $X$ expansion 
 of  $Z^{\reptrip}_{\rmS,4}$  for the   increasing  $N$
 (the coefficients that are stable under the increase of $N$   are  in bold face)
 \be
 \la{finN1}
 \begin{array}{ll}
 \toprule
 N & \qquad\qquad\qquad Z^{\reptrip}_{\rmS,4} \\
 \midrule
 2 & 1+5\,x^{2}+40\,x^{3}+212\,x^{4}+1080\,x^{5}+6054\,x^{6}+\cdots \\
 3 & 1+\bm 6\,x^{2}+\bm{48}\,x^{3}+342\,x^{4}
+2688\,x^{5}+21408\,x^{6}+\cdots\\
 4 & 1+\bm 6\,x^{2}+\bm{48}\,x^{3}+387\,x^{4}+3384\,x^{5}+31765\,x^{6}+\cdots \\
 5 & 1+\bm 6\,x^{2}+\bm{48}\,x^{3}+\bm{396}\,x^{4}+\bm{3504}\,x^{5}+35012
 \,x^{6}+\cdots\\
 6 & 1+\bm 6\,x^{2}+\bm{48}\,x^{3}+\bm{396}\,x^{4}+\bm{3504}\,x^{5}+35535
 \,x^{6}+\cdots\\
 7 & 1+\bm 6\,x^{2}+\bm{48}\,x^{3}+\bm{396}\,x^{4}+\bm{3504}\,x^{5}+\bm{35580}
 \,x^{6}+\cdots\\
 \bottomrule
 \end{array}
 \ee
 The expansions in (\ref{finN1}) may be  derived using the character expansion method that is 
 quite convenient at relatively small $N$
 (see,   \eg, \cite{Balantekin:1983km}). Irreducible representations of $U(N)$ may be labeled by 
 $N$ integers $\bm{n} = (n_{1}, n_{2}, \dots, n_{N})$ with $n_{1}\ge n_{2}\ge \cdots\ge n_{N}\ge 0$.
 Denoting by $t_{1}, \dots, t_{N}$ the eigenvalues of the group element $U$ in the 
 fundamental representation, we obtain the character $\chi_{\bm n}$ from the  Weyl formula
  ($i,j$ are row and column 
 indices)
 \be
 \chi_{\bm n}(U) = \frac{\det(t_{i}^{n_{j}+N-j})}{\det(t_{j}^{N-i})} \ . 
 \ee
 This is a polynomial in the eigenvalues  $t_i$ with total degree equal to 
 $\sum_{i}n_{i}$. Any polynomial 
 built out of  powers of traces $\text{tr}(U^{k})$ may be expanded as a finite sum of  such characters.
 Then  the group integrals in (\ref{3.4}) are easily  evaluated  by exploiting the orthonormality of the characters
$ \int dU\,\chi_{\bm n}(U)\,\overline{\chi_{\bm n'}(U)} = \delta_{\bm n, \bm n'}$. The dependence
on $N$ of the final result is a consequence of the fact that 
the fine details of the character decomposition also depend on $N$. 
To give a simple nontrivial example, let us  consider the integral 
\be
I_{N} = \int dU\, |(\text{tr}\,U^{2})^{2}|^{2}, \qquad U\in U(N),\la{fin33}
\ee
The character expansion of 
$(\text{tr}U^{2})^{2} = (\sum_{i=1}^{N}t_{i}^{2})^{2}$ reads 
\be
 \la{finN2}
 \begin{array}{ll}
 \toprule
 N &  (\text{tr}\, U^{2})^{2}\\
 \midrule
 2 &  2\,\chi_{22}-\chi_{31}+\chi_{40}\\
 3 & 2\,\chi_{220}-\chi_{310}+\chi_{400}-\chi_{211} \\
 4 &  2\,\chi_{2200}-\chi_{3100}+\chi_{4000}-\chi_{2110}+\chi_{1111}\\
 5 &  2\,\chi_{22000}-\chi_{31000}+\chi_{40000}-\chi_{21100}+\chi_{11110}\\
 6 &  2\,\chi_{220000}-\chi_{310000}+\chi_{400000}-\chi_{211000}+\chi_{111100}\\
 \bottomrule
 \end{array}
 \ee
and so on. Thus the  cases $N=2,3$ are special, but  there is a stable  pattern
 for $N\ge 4$. Using the orthogonality of the 
characters  and  (\ref{finN2})  we find for \rf{fin33}
\be
I_{2} = 2^{2}+2\times 1^{2} = 6, \qquad I_{3} = 2^{2}+3\times 1^{2} = 7, \qquad I_{N\ge 4} = 2^{2}+4\times 1^{2}=8.
\ee

\section{Details  of analysis of large $N$  partition function for 3-plet  representation}

\subsection{$U(2)$ case}
\la{app:L}

\begin{figure}[t]  
  \centering
 \includegraphics[scale=0.39]{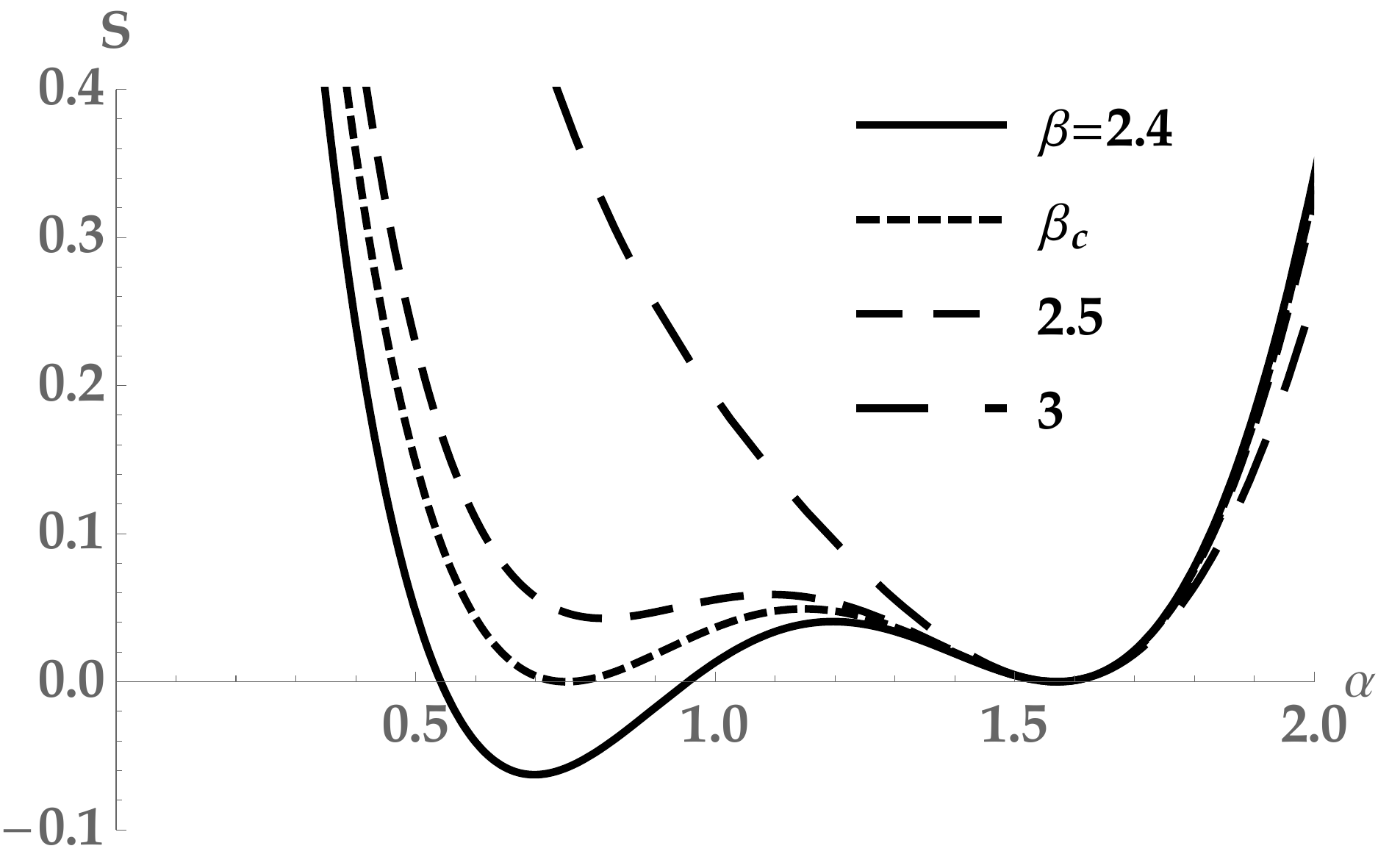}
\includegraphics[scale=0.39]{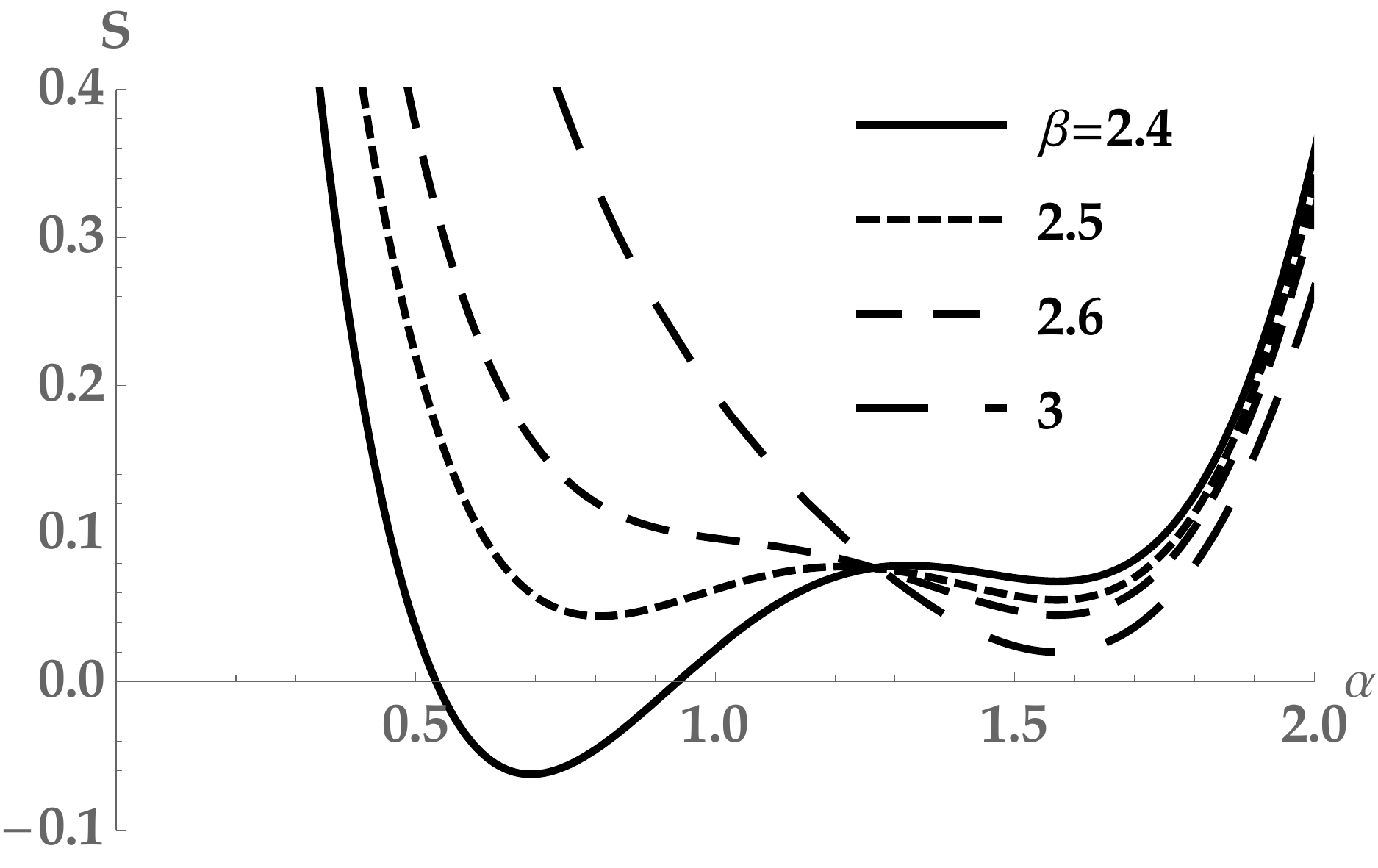}
   \caption[]
   {
   The action  $S$ in (\ref{4.3}) for the $U(2)$ theory as a function of the eigenvalue 
   $\alpha=\alpha_{1}=-\alpha_{2}$.  The left   plot is 
   for one   and the right plot is for ten  harmonics included in \rf{4.3}. 
    \label{fig3}  }
\end{figure}

To compare the  large $N$  and the finite $N$ cases, it   is useful  to consider 
the  lowest  non-trivial value of  $N=2$. Then  the action (\ref{4.3}) in 3-plet case with 
just one harmonic  included is a function of  the single eigenvalue 
angle $\alpha=\alpha_{1} = -\alpha_{2}$  and  inverse temperature $\b$
\be
S = -\log\sin^{2}\alpha-16\,z_\Phi(e^{-\beta})\,\cos^{3}\alpha\ .
\ee
The left part of    Fig.~\ref{fig3} is the    plot of this function for four values of $\beta$.
 For 
$\beta < \beta_{c}$  the value of $S$   at  the global minimum   is   negative.
This global minimum  on the left    and the  local   minimum  at $\pi/2$ on the right 
become degenerate at  $\beta_{c}$. 
 For $\beta>\beta_{c}$ the global
minimum is  on the right at $\alpha=\pi/2$. 

The transition temperature  can be found analytically to be  
$\beta_{c}=2.454$. The associated eigenvalue is $\alpha_{c} = 0.752$. 
This means that there is a first order (discontinuous) transition. Increasing
$\beta$, the eigenvalue $\alpha$ goes from 0 to $\alpha_{c}$ and then jumps to  $\alpha=\pi/2$.
Increasing the number of harmonics included in \rf{4.3}
does not change this  picture qualitatively. This is illustrated in the  right plot in 
 Fig.~\ref{fig3}  where we assumed that  there are 10 harmonics in \rf{4.3}.
 
  What changes at higher $N$ is that the right minimum
shifts  further to the right,  tending to $\alpha=\pi$ for $N\to \infty$.

\subsection{One-harmonic solution:   value of the  action  for  the eigenvalue  density  }
\la{app:K}
To determine the leading  term in  the free energy  $\log Z$  at the  large $N$ saddle point   one is to compute the value of the action  \rf{4.14} on the solution of \rf{5.2}. 

Below   we shall  compute the  sum of the  two terms in the action \rf{4.14}
\begin{align}
S_{M} &= -\frac{1}{2}N^{2} \int d\alpha\,d\alpha'\,\rho(\alpha)\,\rho(\alpha') \te
\,\log\sin^{2}\frac{\alpha-\alpha'}{2}, \\
V &= -2\,N^{3}\,z_\Phi(e^{-\beta})\,\int d\alpha\,d\alpha'\,d\alpha''\,\rho(\alpha)\,\rho(\alpha')\,
\rho(\alpha'')\,\cos(\alpha+\alpha'+\alpha'') 
\end{align}
on the solution (\ref{5.7})  found in one-harmonic approximation, {\em i.e.}
\be
\rho(\alpha) =  \frac{1}{u\,\pi}\,\sqrt{u-\sin^{2}\frac{\alpha}{2}}\,
\cos\frac{\alpha}{2}, \qquad \qquad u = \sin^{2}\frac{\alpha_{0}}{2}.
\ee
Using  the identity
\be
-\frac{1}{2}\log\sin^{2}\frac{\alpha}{2} = \log 2 + \sum_{m=1}^{\infty} \frac{1}{m}\cos(m\,\alpha),
\ee
we find for the measure term 
\be
\la{D.5}
S_{M} = -\frac{1}{2}N^2  \int d\alpha\,d\alpha'\,\rho(\alpha)\,\rho(\alpha')
\,\log\sin^{2}\tfrac{\alpha-\alpha'}{2} = N^{2}\,\Big(\log 2+\sum_{m=1}^{\infty}
\frac{1}{m}\,\rho_{m}^{2}
\Big),
\ee
where $\rho_{m}$ was  defined in (\ref{5.4}), {\em i.e.}
\be\la{d7}
\rho_{m} = \frac{1}{\pi\,u}\,\int_{-\alpha_{0}}^{\alpha_{0}}d\alpha \sqrt{u-\sin^{2}\frac{\alpha}{2}}\,
\cos\frac{\alpha}{2}\,\cos(m\alpha),\qquad u=\sin^{2}\frac{\alpha_{0}}{2}\ . 
\ee
Introducing $t= \sin\frac{\alpha}{2} /\sin\frac{\alpha_{0}}{2}$  we get 
\be
\la{D.7}
\rho_{m} = \frac{2}{\pi\,\sqrt{u}}\,\int_{0}^{1}dt\,\sqrt{1-t^{2}}\sqrt{1-u\,t^{2}}\,
\cos\big[2\,m\,\arcsin(u\,t)\big].
\ee
Using the expansion
\be
\cos\big[2\,m\,\arcsin(u\,t)\big] = \sum_{n=0}^{\infty}\frac{(m)_{n}(-m)_{n}}{(2n)!}\,(2\,u\,t)^{2n},
\ee
we obtain from (\ref{D.7})\footnote{The first cases are
\notag
$
\rho_{0}=1, \quad \rho_{1}=1-\frac{u}{2}, \quad \rho_{2} = 1-2\,u+u^{2}, \quad
\rho_{3} = 1-\frac{9 u}{2}+6u^{2} -\frac{5 u^3}{2}, ...$
Here $\rho_{0}=1$ is the normalization of $\rho$, while $\rho_{1}$ is consistent with (\ref{5.10}).}
\be
\rho_{m} = \sum_{n=0}^m \frac{(m)_{n}\,(-m)_{n}}{n!\,(n+1)!}\,u^{n} = 
{}_{2}F_{1}(-m,m,2; u).
\ee
Thus  
\be
\la{D.55}
S_{M} =  N^{2}\,\Big(\log 2+\sum_{m=1}^{\infty}
\frac{1}{m}\,  \big[{}_{2}F_{1}(-m,m,2; u) \big]^{2}
\Big) \ . 
\ee
The potential term is  given  simply  by 
\be
\la{D.11}
V = -2\,N^{3}\,z_\Phi(e^{-\beta})\,\rho_{1}^{3} = -2\,N^{3}\,z_\Phi(e^{-\beta})\,\big(1-\tfrac{u}{2}\big)^{3} \equiv N^2  \bar V (u)  \ , 
\ee
where   we used \rf{5.10},\rf{5.11}. 

\begin{figure}[t]  
  \centering
 \includegraphics[scale=0.5]{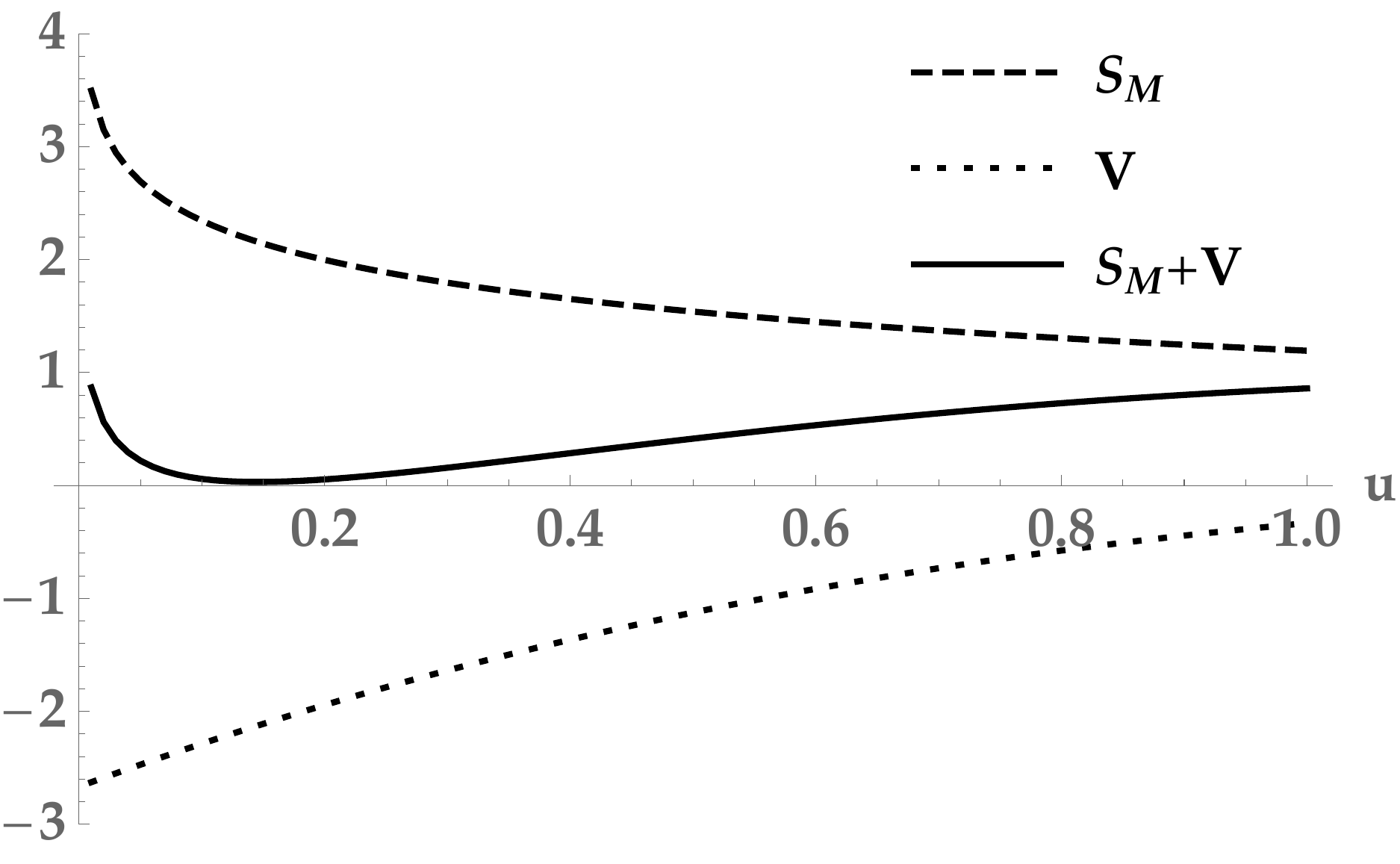}
   \caption[]
   {Plot of the  measure term $S_M$ in   (\ref{D.5}) and the potential term $V$ in  (\ref{D.11}) in the eigenvalue density action 
    at 
   $a_{1}= 3\,N\,z_\Phi(e^{-\beta})=4$. The black line is the total action  that has a minimum at the 
   position predicted by (\ref{5.11}), {\em i.e.} at $u\simeq 0.145$.
    \label{fig2}  }
\end{figure}

In Fig. \ref{fig2}  we plot  $S_{M}$ and $V$ evaluated as functions of $u$
at $a_{1} = 3\,N\,z_\Phi(e^{-\beta})=4$, \ie at the value which is  above  the bound in (\ref{5.13}). As expected, there is a 
minimum of the total action $S= S_M + V$ 
 located at the value predicted by  the  cubic equation for $u$ in (\ref{5.11}).

\subsection{
Including higher harmonics}
\la{app:Q}

It is  easy to  generalize the discussion  in   section 4.4.2
to the case of  higher harmonics  included  in \rf{5.3}.  Some analytical information may be obtained at least
for small $\beta$ where  the maximal   value  of eigenvalues   $\alpha_{0}$ is small. 
We need to solve  eq. \rf{5.2}, i.e. 
\be\!\!\!\!
\la{E.1}
\int d\alpha' \,\rho(\alpha')\,\cot\tfrac{\alpha-\alpha'}{2}=6N\sum_{m=1}^{\infty}\, \zp(x^m) 
\int d\alpha'\,d\alpha''\,\rho(\alpha')\,\rho(\alpha'')\,\sin(m\,(\alpha+\alpha'+\alpha'')).
\ee
Introducing 
\be
\theta ={\alpha\ov \alpha_{0}}\ , \qquad \widetilde\rho(\theta) = \alpha_{0}\,\rho
(\alpha_{0}\,\theta)\ ,
\qquad 
\la{E.3}
d\alpha\,\rho(\alpha) = d\theta\,\widetilde\rho(\theta)\ , \qquad \int_{-1}^{1}d\theta\,\widetilde\rho(\theta)=1
\ee
and expanding (\ref{E.1}) in small $\alpha_{0}$, we obtain at the leading order 
\be
\la{E.4}
\int_{-1}^{1} d\theta'\,\frac{\widetilde\rho(\theta')}{\theta-\theta'}-
3\,N\,\alpha_{0}^{2}\,\sum_{m=1}^{\infty} m\,z_\Phi(e^{-m\,\beta})\,\theta=0\ . 
\ee
For a constant  parameter $\gamma$, the Hilbert problem 
\be
\int_{-1}^{1} d\theta'\,\frac{\widetilde\rho(\theta')}{\theta-\theta'} = \gamma\,\theta,\qquad \qquad \theta\in(-1,1)
\ee
has the unique solution
\be
\widetilde\rho(\theta) = \frac{\gamma}{\pi}\,\sqrt{1-\theta^{2}}.
\ee
The normalization in  (\ref{E.3}) fixes $\gamma=2$. Comparing with (\ref{E.4}), we thus 
 determine
\be
\la{E.7}
\alpha_{0} = \Big[\tfrac{3}{2}\,N\,\sum_{m=1}^{\infty} m\,z_\Phi(e^{-m\,\beta})\Big]^{-1/2}.
\ee
For one-harmonic case, this gives $\alpha_{0} = \big[\frac{3}{2}N\,z_\Phi(e^{-\beta})\big]^{-1/2}$ or 
$u=1/(6\,N\,\zp)+\dots$,  in agreement with (\ref{5.12}).

\subsection{Eigenvalue density for (anti) symmetric 3-plet  representation}
\la{app:F}

We can repeat the analysis of   section \ref{sec4}  
 for the (anti) symmetric 3-plet representation, see (\ref{2.7}).
Here  the action is (\ref{4.3}) with 
\be
{\mc V}^{\pm}(\bm \alpha) = \frac{1}{3}\,\sum_{ijk}\cos(\alpha_{i}+\alpha_{j}+\alpha_{k})
\pm\sum_{ij}\cos(\alpha_{i}+2\,\alpha_{j})+\frac{2}{3}\sum_{i}\cos(3\,\alpha_{i}).
\ee
The stationary-point  equation for the eigenvalues $\alpha_i$ is 
\begin{align}
 \sum_{j\neq i}\cot\tfrac{\alpha_{i}-\alpha_{j}}{2}-\sum^\infty_{m=1} z_\Phi(e^{-m\b})\,\Big[&
\sum_{jk}\sin(m(\alpha_{i} +\alpha_{j}+\alpha_{k}))
\pm\sum_{j}\sin(m(\alpha_{i}+2\,\alpha_{j}))\notag \\
& \pm 2\,\sum_{j}\sin(m(2\,\alpha_{i}+\alpha_{j}))+2\,\sin(3\,m\,\alpha_{i})
\Big]=0.
\end{align}
Written   in  terms of the density of eigenvalues \rf{4.8},  this  becomes 
\begin{align}
\la{F.3}
&\int d\alpha'\,\rho(\alpha')\,\cot\tfrac{\alpha-\alpha'}{2}=\sum^\infty_{m=1}z_\Phi(e^{-m\,\beta})\,\Big[
N\,\int d\alpha'\,\int d\alpha''\,\rho(\alpha')\,\rho(\alpha'')\,
\sin(m(\alpha+\alpha'+\alpha''))\notag \\
& \pm \int d\alpha'\,\rho(\alpha')\,\sin(m(\alpha+2\,\alpha'))
 \pm 2\,\int d\alpha'\,\rho(\alpha')\,\sin(m(2\,\alpha+\alpha'))+\frac{2}{N}\,\sin(3\,m\,\alpha)
\Big].
\end{align}
Comparing to \rf{5.2}  found in the general 3-plet case 
we observe  that  the additional terms appearing in the   (anti) symmetric case are 
 suppressed  at large $N$ by powers of $1/N$.

In more detail, 
in the simple one-harmonic case, we can write (\ref{F.3}) 
in the form of   (\ref{5.3}),\rf{5.4}
\be
\int d\alpha'\,\rho(\alpha')\,\cot\tfrac{\alpha-\alpha'}{2} = 2\,\sum_{m=1}^{3}C_{m}\,\sin(m\,\alpha),
\ee
where 
\be
\la{F.5}
C_{1} = \frac{1}{2}\,(N\,\rho_{1}^{2}\pm \rho_{2})\,\zp, \qquad
C_{2} = \pm\,\zp\,\rho_{1},\qquad
C_{3} = \frac{\zp}{N},
\ee
and then the solution  for the density is   similar to   (\ref{5.5}),(\ref{5.6}). 
Assuming that for $N\to\infty$  with   $N\,\zp$ fixed   we have   $\rho_{1},\rho_{2}$  finite,  
 it is clear  that the effects of (anti) symmetrization are subleading.

\begin{figure}[t]  
  \centering
 \includegraphics[scale=0.5]{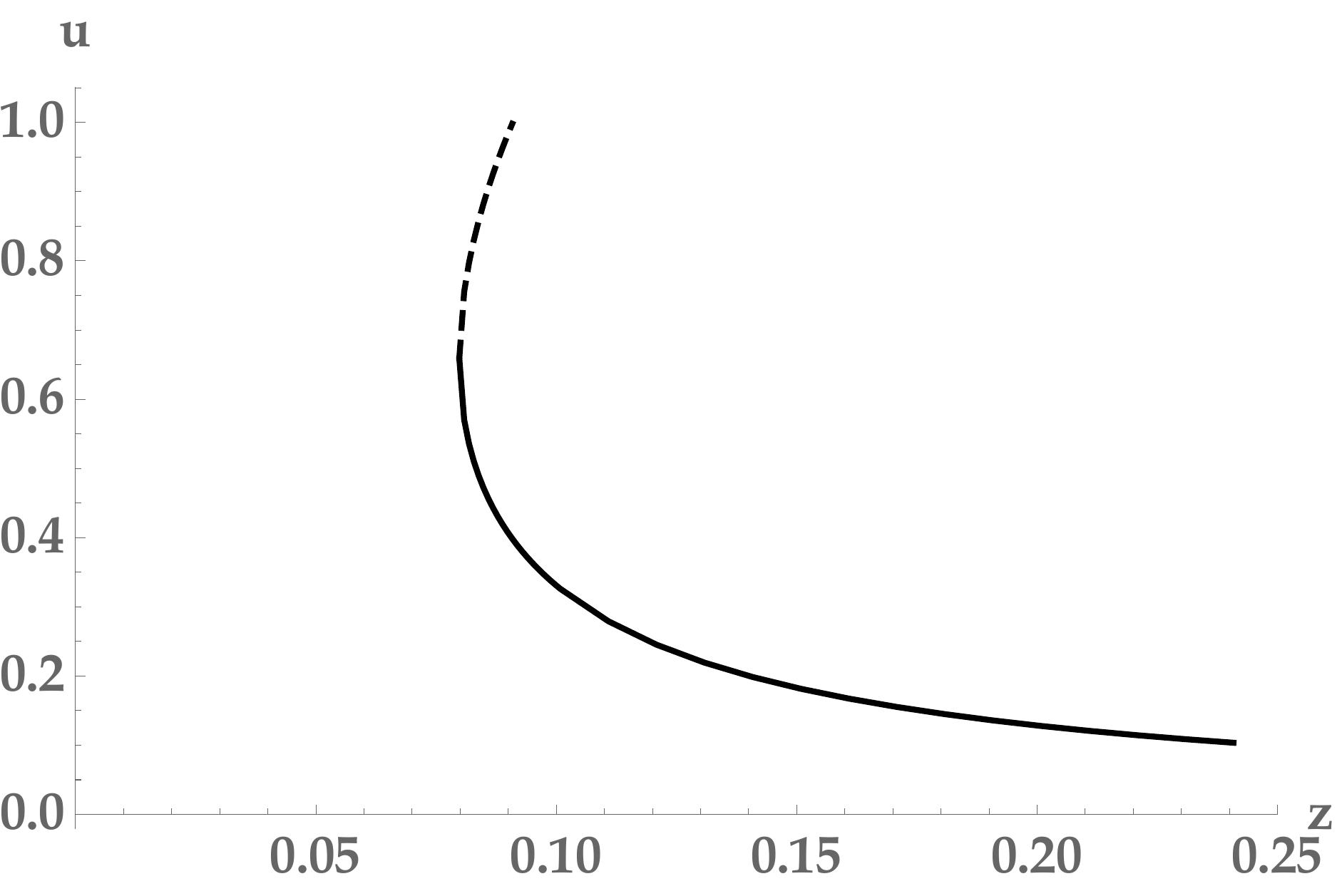}
   \caption[]
   {
    Solution  of the system (\ref{F.7}) 
  for  $u$   as a function of  variable $z\equiv \zp$   for $N=40$. 
        \label{fig4}  }
\end{figure}

 To check these assumptions, let us 
 consider  explicitly the symmetric representation case, {\em i.e.} the plus sign in \rf{F.3},(\ref{F.5}). 
 Introducing  $u=\sin^{2}\frac{\alpha_{0}}{2}$,
the three self-consistency conditions obtained by 
plugging $\rho$ into the definition of $\rho_{1}$ and $\rho_{2}$
and also imposing $\int d\alpha\, \rho(\a)=1$ are
\begin{align}
1&= \frac{2}{N} u \left(10 u^2-12 u+3\right) \zp+N \rho _1^2 u\,  \zp-u\, \zp \left(-4 \rho
   _1-\rho _2+6 \rho _1 u\right) , \notag \\
   \la{F.7}
\rho_{1} &= -\frac{1}{2} N \rho _1^2 (u-2) u \zp-\frac{3}{N} u (5 u-2) (u-1)^2
   \zp  \notag\\
   &\qquad  +\frac{1}{2} u \zp \left(8 \rho _1+2 \rho _2+8 \rho _1 u^2-16 \rho _1
   u-\rho _2 u\right) , \\
\rho_{2}   &= \frac{6}{N} u \left(6 u^2-4 u+1\right) (u-1)^2 \zp+N \rho _1^2 u (u-1)^2 \zp \notag \\
   &\qquad  -u \zp
   \left(-4 \rho _1-\rho _2+9 \rho _1 u^3-20 \rho _1 u^2-\rho _2 u^2+14 \rho
   _1 u+2 \rho _2 u\right) .\notag
\end{align} 
One may study   the properties of the solution $(u,\rho_{1},\rho_{2})$ of the algebraic 
system (\ref{F.7})
for fixed $\zp$ and increasing $N$. A numerical analysis shows that for any $\zp$
 we find exactly one acceptable
solution as soon as $N$ is sufficiently large. This solution may be expanded in powers of $1/N$ and reads
\begin{align}
&\qquad\qquad u = \frac{1}{N \zp} +   \frac{1}{(N \zp)^2}   - {5\ov N} \ \frac{1}{N \zp}   
+\dots, \\ &  \rho_{1} = 1-\frac{1}{2 N \zp} -   \frac{1}{2 (N \zp)^2 } +\frac{5 }{2 N} \ { 1\ov N \zp} 
   +\dots, \\  & \qquad  
\rho_{2} = 1-\frac{1}{2 N \zp} -   \frac{1}{ (N \zp)^2 } +\frac{10 }{ N} \ { 1\ov N \zp} 
   +\dots\ .
\end{align}
Thus the large $N$ behaviour at fixed $N \zp $ is similar to the asymmetric 
 3-plet  case, with 
$\alpha_{0}\sim (N \zp)^{-1/2}$. If we fix $N$ and vary $\zp$, the solution may develop branches
and may exist only in certain ranges. One example is in Fig.~\ref{fig4} where we show the solution 
for $u$  as a function of  $\zp$  for $N=40$. There is a minimal  value of  $\zp$ and also a narrow region 
where both  branches are present. Completely similar features are observed in the case of the 
antisymmetric  3-plet representation.

\section{$N= \infty $ limit  of   low temperature expansion of $O(N)$ partition function} 
\la{app:X}

If the symmetry group is $O(N)$, we   may again start    with 
 the general  expression   for the partition function in (\ref{2.5}). 
 Renaming matrix $U$  as   $M\in O(N)$, the  characters  of the relevant  representations $R$ are 
\be
\la{A.1}
\begin{array}{ccc}
\addlinespace\toprule
R & & \chi_{R} \\
\toprule
{\rm vector} : N & \phantom{\qquad} & \text{tr} (M) \\
{\rm adjoint}: (N\otimes N)_{A} && \frac{1}{2}\,\text{tr}(M)^{2}-\frac{1}{2}\,\text{tr}(M^{2})\\
\text{3-plet} : N^{\otimes 3} && \text{tr}(M)^{3}\\
\bottomrule\addlinespace
\end{array}
\ee
Expanding (\ref{2.5}) in powers of the matrix $M$, we are led to the problem of computing the $O(N)$ 
group integrals  parametrized by  the integers $ \bm{a}= \{ a_\ell\}$

\be
\la{A.2}
I(\bm a) = \int dM\,\prod_{\ell\ge 1} (\text{tr}\, M^{\ell})^{a_{\ell}}\ . \ee
As in the $U(N)$ case in \rf{3.4}, if $N$ is sufficiently large, $I(\bm a)$ does not depend on $N$ and factorizes. 
The precise condition is $N\ge 2\,\kappa(\bm a)$  where $\kappa(\bm a)$ was defined in \rf{344}. 
In this case one can represent the integral in the form 
\be
\la{A.3}
I(\bm a) = \int \prod_{\ell}\frac{d\xi_{\ell}}{\sqrt{2\pi}}\, e^{-\frac{1}{2}\xi_{\ell}^{2}}\ 
 \prod_{\ell\ge 1}(\sqrt{\ell}\,\xi_{\ell}+\eta_{\ell})^{a_{\ell}}\ , 
\qquad\qquad \te  \eta_{l} = \frac{1+(-1)^{\ell}}{2},
\ee
where   $\xi_{\ell}$ are   independent normal variables with Gaussian distribution
\cite{pastur2004moments}.
As a result, 
\be
\la{A.4}
I(\bm a) = \prod_{\ell\ge 1}\begin{cases}
\frac{1+(-1)^{a_{\ell}}}{2}\,(2\,\ell)^{a_{\ell}/2}\,\frac{1}{\sqrt\pi}\Gamma(\frac{a_{\ell}+1}{2}),
&\ \ \  \ell\ \text{odd}, \\
\sum_{n=0}^{a_{\ell}}\binom{a_{\ell}}{n}\, \frac{1+(-1)^{n}}{2}\,(2\,\ell)^{n/2}\,\frac{1}{\sqrt\pi}\Gamma(\frac{n+1}{2}),
&\ \ \  \ell\ \text{even}.
\end{cases}
\ee
Using this  result, 
we may    determine  the low temperature expansion of the 
partition function for a 4d scalar field transforming in various representations like  in  (\ref{A.1}). 
In the vector representation  case  we get  (cf. \rf{3.7})
\be
\la{A.5}
Z^{\repvec}_{\rmS,4}  =1+\,x^2+4 \,x^3+20 \,x^4+56 \,x^5+164 \,x^6+412 \,x^7+1116 \,x^8+
\dots.
\ee
This agrees with the result of  \cite{Giombi:2014yra} for the large $N$  partition function
in the $O(N)$  case which is given by  \rf{333}    with  the following  "single-trace"  partition function 
(cf. \rf{3.1})\foot{Here the  scalar is real so the $x^{2}$ term  corresponds to the operator 
 $\varphi_{i}\varphi_{i}$. The coefficient 4 
of  $x^{3}$ term  comes from  $\varphi_{i}\partial_{\mu}\varphi_{i}$. The 19 $x^{4}$ term
comes from 9 operators $\varphi_{i}\partial_{\mu}\partial_{\nu}\varphi_{i}$ and $\frac{1}{2}(
4\times 5) = 10$ operators
 $\partial_{\mu}\varphi_{i}\partial_{\nu}\varphi_{i}$, etc.}
\begin{align}
\la{A.6}
Z_{\rm s.t.}^{\repvec} &=\te  \frac{1}{2}\,\big[z_\Phi(x)\big]^{2}+\frac{1}{2}\,z_\Phi(x^{2}) \notag \\
&= 
\,x^2+4 \,x^3+19 \,x^4+52 \,x^5+134 \,x^6+280 \,x^7+554 \,x^8+984 \,x^9+\dots.
\end{align}
 In the adjoint scalar  case, we find (cf. \rf{3.8})
 \begin{align}
 Z^{\repadj}_{\rmS,4} &= 1+\,x^2+4 \,x^3+21 \,x^4+66 \,x^5+235 \,x^6+724 \,x^7+2423 \,x^8+7873 \,x^9\notag \\
 & \ \ \quad\  +26463 \,x^{10}+88252
   \,x^{11}+297918 \,x^{12}+1003530 \,x^{13}+\dots, 
\end{align}
with the corresponding  single-trace partition function in \rf{333} being 
\begin{align}
\la{A.8}
Z_{\rm s.t.}^{\repadj} &= \,x^2+4 \,x^3+20 \,x^4+62 \,x^5+204 \,x^6+578 \,x^7+1730 \,x^8\notag \\
&\ \ \ \ \ +5073 \,x^9+15495 \,x^{10}+47791
   \,x^{11}+\dots\ . 
 \end{align}
$Z_{\rm s.t.}^{\repadj}$   counts  the operators which 
 are single  traces of products of  scalars fields   which are antisymmetric  $O(N)$ matrices.
 Here  we have the identity
\be
\la{A.9}
\text{tr}(\Phi_{1}\Phi_{2}\dots\Phi_{n}) = (-1)^{n}\,\text{tr}(\Phi_{n}\Phi_{n-1}\dots\Phi_{1}) \ , 
\ee
where $\Phi_{n}$ is the scalar $\varphi_{ij}=-\varphi_{ji}$ 
or any  derivative of it. It seems non-trivial to apply Polya counting in the case of an additional  constraint (\ref{A.9}), and we did not find 
a simple  closed  formula  for   $  Z_{\rm s.t.}  $ like (\ref{3.2}). To see the non-trivial effect of the constraint (\ref{A.9}), let us explicitly list  the single-trace operators up 
 to dimension 5:   
 {\small 
\be
\begin{array}{ccc}
\toprule
\dim & \text{operator} & \text{multiplicity} \\ \toprule
2 & \text{tr}(\varphi\varphi) & 1 \\
\midrule
3 & \text{tr}(\varphi\partial_{\mu}\varphi) & 4 \\
\midrule
4 & \text{tr}(\varphi\partial_{\mu}\partial_{\nu}\varphi)  & 9 \\
  & \text{tr}(\partial_{\mu}\varphi\partial_{\nu}\varphi)  & 10 \\
  & \text{tr}(\varphi\varphi\varphi\varphi) & 1 \\
\midrule
5 & \text{tr}(\varphi\partial_{\mu}\partial_{\nu}\partial_{\rho}\varphi) & 20-4 = 16\\
 & \text{tr}(\partial_{\mu}\varphi\partial_{\nu}\partial_{\rho}\varphi) & 4\times 9 = 36\\
    & \text{tr}(\varphi\partial_{\mu}\varphi\partial_{\nu}\varphi) & 6 \ (\mu\neq \nu)\\
    & \text{tr}(\varphi\varphi\varphi\partial_{\mu}\varphi) & 4\\
\cmidrule{2-3}
\end{array}
\ee
}
Note that $\text{tr}(\varphi\varphi\partial_{\mu}\partial_{\nu}\varphi) =0$ in view of  (\ref{A.9}).
Also,  $\text{tr}(\varphi\partial_{\mu}\varphi\partial_{\nu}\varphi)$ with $\mu=\nu$
is again  zero in view of  (\ref{A.9}). The resulting  multiplicities 1, 4, 20, 62 are in 
agreement with (\ref{A.8}).

\def \tr  {{\rm tr}}

\medskip
The case   the symmetric representation $(N\otimes N)_{S}$ appears to be simpler.  Here  the analog of  (\ref{A.9})
reads
\be
\la{A.11}
\text{tr}(\Phi_{1}\Phi_{2}\dots\Phi_{n}) = \text{tr}(\Phi_{n}\Phi_{n-1}\dots\Phi_{1}),
\ee
and it adds an extra symmetry to the standard cyclic invariance of the trace. 
Then  the total and single-trace partition functions are found to be 
\begin{align}
Z^{\text{2-plet}^+}_{\rmS,4}  &= 1+\,x+6 \,x^2+20 \,x^3+75 \,x^4+246 \,x^5+862 \,x^6+2852 \,x^7\notag \\
& \qquad +9643 \,x^8+32040 \,x^9+107141 \,x^{10}+356651
   \,x^{11}+1191345 \,x^{12}+\dots, \\
\la{A.13}
Z^{\text{2-plet}^+}_{\rm s.t.} &= \,x+5 \,x^2+14 \,x^3+40 \,x^4+101 \,x^5+276 \,x^6+715 \,x^7
\notag \\
& \qquad +1982 \,x^8+5553 \,x^9+16379 \,x^{10}+49476
   \,x^{11}+154346 \,x^{12}+\dots.
\end{align}
One  can  find a  closed form of  (\ref{A.13})   using  the Polya enumeration theorem  and taking into  account that the symmetry group is the cyclic group with an 
additional inversion (\ref{A.11}). A careful examination of the cycle structure of the associated
permutations  gives
\begin{align}
\la{A.14}
Z^{\text{2-plet}^+}_{\rm s.t.} = & -\frac{1}{2}\,\sum_{m=1}^{\infty}\frac{\varphi(m)}{m}\,
\log\big(1-z_\Phi(x^{m})\big)\notag \\
& +\frac{1}{2}\,\sum_{m=1}^{\infty}\sum_{\ell=1}^{k}\frac{1}{m}\,\begin{cases}
\begin{cases}
\big[z_\Phi(x^{2})\big]^{\frac{m}{2}} & \ell\ \text{even} \\
\big[z_\Phi(x^{2})\big]^{\frac{m}{2}-1}\,\big[z_\Phi(x)\big]^{2} & \ell\ \text{odd}
\end{cases} & \qquad m\ \text{even} \\
\big[z_\Phi(x^{2})\big]^{\frac{m-1}{2}}\,z_\Phi(x) &\qquad  m\ \text{odd}.
\end{cases}
\end{align}
where $\varphi(m)$ is the same as in \rf{3.2}  and  the additional $1/2$ factors are
 due to the fact that the symmetry group for a trace with $m$
objects is $2m$ (from $m$ shifts and  $m$ reflected shifts).\footnote{The  presence of extra  terms 
in the second line of  (\ref{A.14})
is due to the fact 
that a reflected shift by $\ell$ places  of  a string of $m$ objects splits into: 
(i) $m\ov 2$ \  2-cycles if $\ell, q$ are even; (ii) $m-2\ov 2$\  2-cycles and 2 1-cycles if $\ell$ is odd and $m$
is even; (iii) $m-1\ov 2$ \ 2-cycles and one 1-cycle if $m$ is odd.}  

Using  (\ref{A.14}),
we have computed the series expansion (\ref{A.13}) up to the very
 high order $\mc O(x^{100})$
and a numerical analysis revealed that the series is convergent  for $x < x_c$ with 
$z_\Phi(x_c)=1$ (for the 4d scalar this  critical value is $x_c\simeq 0.285$). 
The same  behaviour was found in the $U(N)$ case so the conclusion is that 
  the additional terms in 
 the second line of (\ref{A.14})
do not worsen the  convergence. 

In  the 3-plet case representation case we obtain  (cf. \rf{3.7})
\begin{align}
Z^\reptrip_{\rmS, 4} &= 1+11 \,x^2+60 \,x^3+773 \,x^4+7920 \,x^5+110781 \,x^6+1509060 \,x^7\notag \\
 &\qquad  +23807838 \,x^8+379566780
   \,x^9+6645202174 \,x^{10}+118587559020 \,x^{11}\notag \\
   &\qquad +2264713625957 \,x^{12}+44204970285420
   \,x^{13}+\dots, \la{A.15}
\end{align}
with  the corresponding  "single-trace" partition function being 
\begin{align}
\la{A.16}
Z_{\rm s.t.}^\reptrip = &11 \,x^2+60 \,x^3+707 \,x^4+7260 \,x^5+100888 
\,x^6+1382820 \,x^7+21944399 \,x^8\notag \\
& +352168900 \,x^9+6207336278
   \,x^{10}+111438968700 \,x^{11}+\dots.
 \end{align}
To reproduce the  $x^2$ term here   by counting dimension 2   operators 
we need to  classify   various   bilinear contractions:   (i)  there are $\frac{1}{2}(3\times 4) = 6$  contractions containing traces 
$
\varphi_{iij}\varphi_{iij}, \ \varphi_{iij}\varphi_{iji} ,\ \dots,
$
where  we need to account that position of the index  contracted between the two fields matters  and that there is a symmetry between  the two fields in the real   scalar case;
(ii) there are also $3! $  irreducible  contractions 
$\varphi_{ijk}\varphi_{ijk}, \ \  \varphi_{ijk}\varphi_{ikj},\   \dots,
$
but one needs to take into account the symmetry relation
$\varphi_{ijk}\,\varphi_{jki} \equiv \varphi_{ijk}\,\varphi_{kij}$, so  we are left with $6-1=5$
independent choices. The total $6+5=11$ matches the coefficient of the $x^2$ term in 
\rf{A.15},\rf{A.16}.

There are  fewer  operators in the case of 
totally symmetric or  antisymmetric  3-plet representations, i.e.   the coefficients
in the small $x$ expansion of $Z$   should  be much smaller. Indeed, we find
directly from \rf{2.7}   (cf. \rf{3.17},\rf{3.18})
\begin{align}
Z^{\reptrip^{+}}_{\rmS, 4}  = 1+&2 \,x^2+8 \,x^3+46 \,x^4+156 \,x^5+668 \,x^6
+2684 \,x^7+12044 \,x^8\notag\\
& +53556 \,x^9+249495 \,x^{10}+1182524
   \,x^{11}+5780012 \,x^{12}+\dots, \\
Z^{\reptrip^{+}}_{\rm s.t.} =& 2 \,x^2+8 \,x^3+43 \,x^4+140 \,x^5+542 \,x^6+2036 \,x^7
+8688 \,x^8\notag \\
& +37868 \,x^9+175063 \,x^{10}+832556
   \,x^{11}+4102153 \,x^{12}+\dots, 
\\
Z^{\reptrip^{-}}_{\rmS, 4}  =1+&\,x^2+4 \,x^3+22 \,x^4+64 \,x^5+228 \,x^6+716 \,x^7+2701 \,x^8
\notag \\
&+10104 \,x^9+41897 \,x^{10}+179420
   \,x^{11}+820244 \,x^{12}+\dots,  \\
Z^{\reptrip^{-}}_{\rm s.t.} = &\,x^2+4 \,x^3+21 \,x^4+60 \,x^5+196 \,x^6+568 \,x^7+2002 \,x^8
\notag \\
& +7324 \,x^9+30768 \,x^{10}+136096 \,x^{11}+644817   \,x^{12}+\dots.
\end{align}
For  example, the first  few  single-trace states in the antisymmetric case are 
\be
\begin{array}{ccc}
\toprule
\dim & \text{operator} & \text{multiplicity}\\ \toprule
2 & \varphi_{ijk}\,\varphi_{ijk} & 1 \\
\midrule
3 & \varphi_{ijk}\,\partial_{\mu}\varphi_{ijk} & 4 \\
\midrule
4 & \varphi_{ijk}\,\partial_{\mu}\partial_{\nu}\varphi_{ijk} & 9 \\
  & \partial_{\mu}\varphi_{ijk}\,\partial_{\nu}\varphi_{ijk}  & 10 \\
  & \varphi_{ijk}\,\varphi_{ijl}\varphi_{pqk}\varphi_{pql} & 1 \\
    & \varphi_{ijk}\,\varphi_{ipq}\varphi_{jpl}\varphi_{kql} & 1 \\
  \bottomrule
\end{array}
\ee
As a final remark, we note  that using the discussion in  \cite{Giombi:2014yra} providing the 
suitable modification of the measure term in (\ref{4.1}) for the $O(N)$ case, it is possible also to 
study  the large $N$ thermodynamics and the structure of phase transitions in this case, 
 but there should be no  qualitative changes  compared to
$U(N)$ case  analyzed   in   section 4.


\section{General expression for single-trace partition function}  
\la{app:II}

Given the $N=\infty$ partition function $Z(x)$   one  can  invert the relation \rf{333}
and find   the single-trace partition function $Z_{\rm s.t.}(x)$   that counts
only  irreducible ("single-trace")  contractions  among all  singlet operators.
This was  already discussed  in  \ci{Gibbons:2006ij}  and here we present an equivalent but  slightly  more explicit 
 version of this inverse  relation. 

Starting with the relation 
\be
\la{xxx1}
L(x) = \sum_{m=1}^{\infty}\frac{1}{m}\,Z_{\rm s.t.}(x^{m}), \qquad\qquad  L(x) \equiv  \log Z(x)\ , 
\ee
we find 
\be
\la{xxx2}
Z_{\rm s.t.}(x) = L(x)+\sum_{m\in \Omega}(-1)^{\nu_{m}}\,\frac{1}{m}\,L(x^{m}) \ . 
\ee
Here $\Omega = \{2,3,5,6,7,10,11,13,\dots\}$ is the set of so-called square-free 
integers,  such that their prime number factorization is of the form 
$m=\prod_{i=1}^{\nu_{m}} p_{i}$,
{\em i.e.}  is  the product of 
 prime factors each appearing  in the  first power only. 
 The sign  factor 
$(-1)^{\nu_{m}}$ is known   in this context 
 as the Liouville function. 
 The proof of (\ref{xxx2})  is by   substituting   \rf{xxx2} into  (\ref{xxx1}):
\begin{align}
\sum_{m=1}^{\infty}\frac{1}{m}\,Z_{\rm s.t.}(x^{m}) &= \sum_{m=1}^{\infty}\frac{1}{m}\,
L(x^{m})+\sum_{n=1}^{\infty}\sum_{q\in\Omega}(-1)^{\nu_{q}}\frac{1}{n\,q}L(x^{nq}) \notag \\
&= 
 \sum_{m=1}^{\infty}\frac{1}{m}\,
L(x^{m})\Big[1+\mathop{\sum_{q\in\Omega}}_{q|m}(-1)^{\nu_{q}}\Big] = L(x) \ . 
\end{align}
In the last equality, we used  that 
\be
\la{xxx3}
\mathop{\sum_{q\in\Omega}}_{q|m}(-1)^{\nu_{q}} = \begin{cases}
1\ ,  & \ \ m=1, \\
0\ ,  & \ \ m>1.
\end{cases}
\ee
To prove (\ref{xxx3}), it is enough to observe that if $m>1$ is factorizable into
  $M$ powers of distinct primes,  $m= p_{1}^{a_{1}}\cdots p_{M}^{a_{M}}$, 
then 
\be
1+\mathop{\sum_{q\in\Omega}}_{q|m}(-1)^{\nu_{q}} ={\te 1-M+\frac{M(M-1)}{2}-\frac{M(M-1)(M-2)}{3!}}
+... = (1-1)^{M}=0,
\ee
where we used that  the square-free integers $q$ dividing $m$ are of the form 
$p_{i}$, $p_{i}p_{j}	\ (i\neq j)$, {\em etc.}

Let us note that if $L(x)$ in \rf{xxx1}   has the form 
\be
L(x) = \sum_{m=1}^{\infty}f(x^{m})\ ,
\ee
where  $f(x) $ is  independent of $m$, then one can show that (\ref{xxx2}) gives
\be
Z_{\rm s.t.}(x) = \sum_{m=1}^{\infty}\frac{\varphi(m)}{m}\,f(x^{m})\ , 
\ee
where $\varphi$   is the Euler's totient function as in \rf{3.2}. 
One example is the  adjoint  representation case in \rf{3.3}. 
Another is the 2-plet  representation with $Z$ given in \rf{B.3}   in which case   (cf. \rf{3.2})
\be \la{E.8}
Z^{\text{2-plet}}_{\rm s.t.}(x) = -{1\ov 2} 
\sum_{m=1}^{\infty}\frac{\varphi(m)}{m}\,\log\big(1- 4\big[\,z_{\Phi}(x^{m})\big]^2\big)\ .
\ee
 

\def \tens{p\text{-tensor}}

\section{
  Singlet  partition function of $[U(N)]^p$ invariant $p$-tensor theory}  
\la{app:FF}

Given a  free $p$-tensor $\Phi=(\vp_{i_1...i_p})$   with each internal index running from from 1 to $N$  one 
 may  have several options  of 
  how to define the  corresponding  CFT     and thus  the associated  singlet partition function $Z(x)$
  (i.e. which singlet operators 
  given by   contractions   of fields   to include). 
  In the main part of this  paper we treated all $p$  indices of  $(\vp_{i_1...i_p})$ 
   as equivalent and  thus  all of their contractions were allowed. 
  The  corresponding singlet  partition function on $S^1_\b \times S^{d-1}$   was  then   found by  gauging the 
  global $U(N)$  or $O(N)$ symmetry. 
 
   If  instead all $p$ indices are   assumed to be distinguishable as, e.g.,  in the  interacting  tensor models  considered in \ci{Witten:2016iux,Klebanov:2016xxf}, 
   then the   singlet constraint   may  be implemented by gauging the  
   full $[U(N)]^p$  symmetry  group  \ci{Klebanov:2016xxf}.
     As we shall 
   demonstrate below,   in this  case    the    low temperature  expansion of $Z(x)$  will  again 
   diverge  in the $N=\infty$ limit  
   starting with the $p=3$ case, i.e. the critical temperature  will 
    vanish with $N \to \infty$ if $p \geq 3$.\foot{We are thank  I. Klebanov and 
   G. Tarnopolsky  for  the suggestion   to investigate this case.} 
 
 The computation of the  singlet  partition   function  in the $[U(N)]^p$ invariant theory  (that we will call "$p$-tensor" theory for short) 
 turns out to be very similar to the case of the  $U(N)$ invariant $p$-plet   theory considered in sections 3 and 4    above. 
 To compute $Z$  we may start with the  path integral for $\vp_{i_1...i_p}$  with covariant   derivative   containing $p$ independent 
flat gauge fields   $A_\mu^{(r)}$  ($r=1, ..., p$)   and    average over their $S^1$ holonomies, i.e. constant  $N \times N$ 
hermitian $A_0^{(r)}$   matrices   with the eigenvalues   $\a_{ir}$, or, equivalently, over $U(N)$  matrices $U_r$ 
with  eigenvalues $e^{i \a_{ir}}$  ($i=1, ...., N; \ r=1, ..., p$). 
 As  $\vp_{i_1...i_p}$   transforms in the direct product of fundamental 
representations of $p$  copies of   $U(N)$ group,  the  potentials $A_0^{(r)}$    or the eigenvalues   $\a_{ir}$ 
simply sum up,  i.e. the  resulting  partition function will be a straightforward generalization 
of \rf{2.5}   for the 
  fundamental   representation of a single $U(N)$
    with  the character  of the real representation $R$ in \rf{2.5} now being 
\be
\la{ff0}
\chi(U_{1}, \dots, U_{p}) = \prod_{r=1}^{p}\text{tr}\,  U_{r}+
\prod_{r=1}^{p}\text{tr}\,  U_{r}^{-1}.
\ee
Considering the case of   low temperature expansion of $Z$ in the $N=\infty$ limit 
one finds, doing $p$  independent $dU_{1}\cdots dU_{p}$  integrations    in the same way  as in 
$U(N)$ $p$-plet case in \rf{mb1},\rf{mb2},  
\begin{align}
\la{ff3}
Z^{\tens} 
&= 
\prod_{m=1}^{\infty}\mathop{\sum_{k=0}}^{\infty}
\frac{1}{(2\,k)!}\,\Big(\frac{z_{\Phi}(x^{m})}{m}
\Big)^{2\,k}\binom{2\,k}{k}\,m^{p\,k}\,(k!)^{p} 
= \prod_{m=1}^{\infty}G_{p}\big(m^{p-2}\,\big[z_{\Phi}(x^{m})\big]^{2}\big)\ ,
\end{align}
where  here the  role of $F_p$ in  \rf{mb2},\rf{c1}    is played by the power series 
\be
\la{ff4}
G_{p}(y) = \sum_{k=0}^{\infty} g_k\,y^{k} \ , \ \ \ \ \quad  \ \ \ \ \ \   g_k= (k!)^{p-2} \ , \ \ \quad     p=1,2,3, ...  \ . 
\ee
As \rf{ff3}   involves  the square  of $z_{\Phi}$  and thus is not sensitive to sign  factor in \rf{fff}
it looks the same   for  both pure   boson  or  pure fermion cases. 

The $p=1$   is  of course the standard vector  or 1-plet  case when  $G_{1}(y) = F_1(y) =e^{y}$ as in \rf{c3},\rf{c4}.
In the 2-tensor case  we get 
$G_{2}(y) = 1/(1-y)$  (cf. \rf{c3}) and thus 
\be
Z^{2\text{-tensor}} = \prod_{m=1}^{\infty}\big (1-\big[\zp(x^{m})\big]^{2}\big)^{-1}.
\ee
This is   similar to the adjoint  $U(N)$ case  \rf{3.3} (with $\zp\to \zp^{2}$)\foot{This relation  can be understood
in terms of counting operators as follows:
each singlet can be
considered as built out of  elementary fields    $P_{ij}=\overline\varphi_{ik}\varphi_{jk}$ (with possible 
derivatives)
 with these $P$-fields  contracted in a matrix-like  style.  
Alternatively,  one may  build  all  singlets using  the basis  of 
$P'_{ij}=\overline\varphi_{ki}\varphi_{kj}$. 
} 
 and also to the 2-plet case  \rf{c5},\rf{B.3}.
 
 As in   the $p$-plet case in \rf{c1}, the $p=3$ is the critical value: 
since  $g_{k+1}/g_k = (k+1)^{p-2} $  the series  $G_p(y)$  in \rf{ff4}    does not converge for $p\geq3$. 
The function \be 
G_{3}(y)=\sum_{k=0}^{\infty}k!\,y^{k}\ee 
  that has zero radius of convergence
can be Borel-resummed 
for $y<0$   giving   (cf. \rf{cc5})  
\be
\widetilde G^{B}_3 (y) = -  y^{-1} \,e^{-y^{-1}}\,\Gamma\big(0,-y^{-1}\big) \ ,
\ee
where $ \Gamma(s,y) = \int_y^{\infty} dt\, e^{-t}\,  t^{s-1}\, $ is the incomplete $\Gamma$ function.
Thus  
 $G_3(y)$   is an asymptotic expansion of $\widetilde G^{B}_3 (y)$   for $y<0$. 
 $\widetilde G_{B}(y >0)$ has an
  imaginary part  $\frac{\pi}{y}\,e^{-1/y}$  that   vanishes  exponentially fast for $y\to 0^{+}$.

For example,  in the case of  a  3-tensor  field  $\Phi$ being a  4d scalar  we find from \rf{ff3},\rf{ff4} 
\begin{align}
Z^{3\text{-tensor}}_{\rmS,4} = 1 &+\,x^2+8 \,x^3+38 \,x^4+136 \,x^5+550 \,x^6\notag \\ 
&+2224 \,x^7+9727 \,x^8+42592 \,x^9 +191836 \,x^{10}
+\dots. \la{ff5}
\end{align}
Comparing this to the 3-plet case in \rf{3.9}   we see much smaller coefficients, i.e. 
 the number of  singlet operators is reduced at each   order in dimension.\footnote{Following the remark 
 in   footnote \ref{ft13}, one can evaluate \rf{ff3}  with $\zp(x)\to x $  
corresponding to  partition function  of a constant   scalar field in 4d.
 The  resulting  analog (or "truncation")  of  \rf{ff5} 
 will be  a series in $x^2$ with the  coefficients  given by the  known 
integer sequence A110143  \url{http://oeis.org/A110143}, see also eq.~(19) of \cite{Geloun:2013kta}.
 These coefficients  have asymptotic factorial growth implying again zero radius of convergence. 
}

The single-trace partition function in \rf{333},\rf{xxx2}  corresponding to (\ref{ff5}) is 
\be
\la{ff6}
Z_{\rm s.t.}^{3\text{-tensor}}= x^2+8 \,x^3+37 \,x^4+128 \,x^5+476 \,x^6+1792 \,x^7+7450 \,x^8+31704 \,x^9 + \dots\ ,  
\ee
   and the  lowest coefficients  here are reproduced  by the    operator counting as follows: 
\be
\begin{array}{ccc}
\toprule
\dim & \text{operator} & \text{multiplicity}\\ \toprule
2 & \overline\varphi_{ijk}\,\varphi_{ijk} & 1 \\
\midrule
3 & \overline\varphi_{ijk}\,\partial_{\mu}\varphi_{ijk} \ \text{and  c.c.} & 2\times 4 = 8 \\
\midrule
4 & \overline\varphi_{ijk}\,\partial_{\mu}\partial_{\nu}\varphi_{ijk}  \ \text{and  c.c.} & 2\times 9 = 18 \\
  & \partial_{\mu}\overline\varphi_{ijk}\,\partial_{\nu}\varphi_{ijk}  & 4\times 4 = 16 \\
  & \overline\varphi_{ijk}\,\varphi_{irs}\overline\varphi_{lrs}\varphi_{ljk} & 3 \\
  \bottomrule
\end{array}
\ee
where the multiplicity 3  in last row corresponds to the position of  the index contracted between 
$\overline\varphi$ and $\varphi$.

As in the 3-plet case discussed in section \ref{sec4},   the zero radius of convergence of the low 
temperature series   for  the $[U(N)]^3$ singlet  partition function  $Z^{3\text{-tensor}}$ at $N =\infty$ 
should be related to the vanishing of the corresponding critical temperature 
in the limit  $N\to \infty$. This can be seen explicitly by  repeating the analysis in section \ref{sec4}  in the 3-tensor case. 
Here we will have 3 sets of eigenvalues  $\a_{ir}$ ($r=1,2,3$) and thus 3  densities $\r_r(\a)$    with the  analog  of the 
 action \rf{4.14},\rf{4.15},\rf{4.19}    being 
 \begin{align}
&S(\rho_r,x) = N^{2}\, \sum_{r=1}^3 \int d\alpha\,d\alpha'\,K(\alpha-\alpha')\,\rho_r(\alpha)\,\rho_r(\alpha')\notag \\
& \ \ -  2\,N^{3}\,\int d\alpha\,d\alpha'\,d\alpha''\, \rho_1(\alpha)\,\rho_2(\alpha')\,\rho_3(\alpha'')
\,\sum^\infty_{m=1} { 1\ov m} \zp(x^m) \,\cos\big[m\,(\alpha+\alpha'+\alpha'')\big].\la{419}
\end{align}
It is natural  to look for a stationary  point solution  with  the three 
equal   densities $\rho_1=\rho_2=\rho_3\equiv \rho(\a)$  
for the three $U(N)$ groups, thus getting the equation 
\be\la{420}
\int d\alpha' \,\rho(\alpha')\,\cot {\te \frac{\alpha-\alpha'}{2}}
=2N \sum^\infty_{m=1}\,\zp(x^m)
\int d\alpha' d\alpha''\,\rho(\alpha')\,\rho(\alpha'')\,\sin\big[m\,(\alpha+\alpha'+\alpha'')\big].\qquad
\ee
This equation is  the same  as in \rf{5.2} up to a factor of 3 in the  r.h.s.   and thus its analysis 
 is similar, implying  that  the critical temperature should  again  scale with $N$ as 
 $T_c \sim (\log N)^{-1} $.

\bibliography{BT-Biblio}
\bibliographystyle{JHEP}

\end{document}